\documentclass[prd,nofootinbib,showpacs,superscriptaddress, preprint]{revtex4-1}
\usepackage{amsmath,amsfonts,amssymb,graphicx,natbib,bm}
\usepackage{color}
\usepackage{slashed}    
\usepackage[colorlinks,citecolor=blue]{hyperref}
\usepackage{color}
\usepackage{url}
\usepackage{cancel}
\usepackage{slashed}
\usepackage{multirow}
\usepackage{rotating}
\usepackage{braket}
\usepackage{float}
\usepackage{subfigure}

\newcommand{\be}{\begin{eqnarray*}}
\newcommand{\ee}{\end{eqnarray*}}

\newcommand{\bee}{\begin{eqnarray}}
\newcommand{\eee}{\end{eqnarray}}
\newcommand{\beeq}{\begin{equation}}
\newcommand{\eeq}{\end{equation}}

\usepackage{xspace}

\newcommand{\ba}{\begin{array}}
\newcommand{\ea}{\end{array}}
\newcommand{\bd}{\begin{displaymath}}
\newcommand{\ed}{\end{displaymath}}
\newcommand{\besub}{\begin{subequations}}
\newcommand{\eesub}{\end{subequations}}
\newcommand{\bea}{\begin{eqnarray}}
\newcommand{\eea}{\end{eqnarray}}

\def\l{\lambda}

\def\q2 {q^2}

\def\bt{\begin{table}}
\def\et{\end{table}}

\begin{document}
\title{Minimal Two-component Scalar Doublet Dark Matter with Radiative Neutrino Mass}

\author{Debasish Borah}
\email{dborah@iitg.ac.in}
\author{Rishav Roshan}
\email{rishav.roshan@iitg.ac.in}
\author{Arunansu Sil}
\email{asil@iitg.ac.in}
\affiliation{Department of Physics, Indian Institute of Technology Guwahati, Assam-781039, India }

\begin{abstract}
We propose a minimal extension of the Standard Model to accommodate two-component dark matter (DM) and light 
neutrino mass.  The symmetry of the Standard Model is enhanced by an unbroken $\mathbb{Z}_2 \times \mathbb{Z}'_2$ 
such that being odd under each $\mathbb{Z}_2$, there exists one right handed neutrino and one inert 
scalar doublet. Therefore, each of the $\mathbb{Z}_2$ sectors contribute to ($i$) light neutrino masses radiatively 
similar to the scotogenic models while ($ii$) the two neutral CP even scalars present in two additional inert doublets 
play the role of dark matters. 
Focussing on the intermediate range of inert scalar doublet DM scenario: $M_W \leq M_{\rm DM} \lesssim 500 \; {\rm GeV}$, 
where one scalar doublet DM can not satisfy correct relic, we show that this entire range becomes allowed within this 
two-component scalar doublet DM, thanks to the inter-conversion between the two DM candidates in presence of neutrino 
Yukawa couplings with dark sector.  
\end{abstract}
\maketitle

\section{Introduction}
There have been irrefutable amount of evidences in favour of the existence of non-luminous, non-baryonic form of matter 
in the universe, popularly known as dark matter (DM).  The presence of this form of matter has also been supported by 
astrophysical observations like the ones related to galaxy clusters by Fritz Zwicky \cite{Zwicky:1933gu} back in 1933, 
observations of galaxy rotation curves in 1970's \cite{Rubin:1970zza}, the more recent observation of the bullet cluster 
by Chandra observatory \cite{Clowe:2006eq} along with several galaxy survey experiments which map the distribution 
of such matter based on their gravitational lensing effects. There is equally robust evidence from cosmology as well, 
suggesting that around $26\%$ of the present universe's energy density is in the form of dark matter. In terms of density 
parameter $\Omega_{\rm DM}$ and $h = \text{Hubble Parameter}/(100 \;\text{km} ~\text{s}^{-1} 
\text{Mpc}^{-1})$, the present DM abundance is conventionally reported as \cite{Aghanim:2018eyx}:
$\Omega_{\text{DM}} h^2 = 0.120\pm 0.001$
at 68\% CL. 

In spite of these astrophysical and cosmology based evidences, there have been no detection of particle DM at any experiments. 
The direct detection experiments like LUX \cite{Akerib:2016vxi}, PandaX-II \cite{Tan:2016zwf, Cui:2017nnn} and Xenon1T \cite{Aprile:2017iyp, Aprile:2018dbl} have continued to produce null results so far. As the Standard Model (SM) of particle physics 
can not accommodate such a form of matter, several beyond the Standard Model (BSM) proposals have been put forward 
\cite{Taoso:2007qk} out of which the weakly interacting massive particle (WIMP) paradigm is the most widely studied one.
While such interactions are capable of explaining correct relic density of DM, the same interactions can also give rise to 
production of DM particles at the large hadron collider (LHC) \cite{Kahlhoefer:2017dnp}. However, nothing is found so far in these searches also, 
putting strict bounds on DM coupling to the SM particles, particularly quarks. Another detection prospect lies in the indirect 
detection frontier where searches are going on to find excess of antimatter, gamma rays or neutrinos, originating perhaps 
from dark matter annihilations (for stable DM) or decay (for long lived DM). While no convincing DM signal has been observed 
yet, there are tight constraints on DM annihilations into SM particles \cite{Ahnen:2016qkx}, specially the charged ones 
which can finally lead to excess of gamma rays for WIMP type DM.

Though the null results mentioned above have not ruled out all the parameter space for a single particle DM models yet, 
it may be suggestive of presence of a much richer DM sector. The idea may be natural given the fact that the 
visible sector is made up of several generations for single type of particles. There have been several proposals for 
multi-component WIMP dark matter during last few years, some of which can be found in \cite{Cao:2007fy, Zurek:2008qg, Chialva:2012rq, Heeck:2012bz, Biswas:2013nn, Bhattacharya:2013hva, Bian:2013wna, Bian:2014cja, Esch:2014jpa, Karam:2015jta, Karam:2016rsz, DiFranzo:2016uzc, Bhattacharya:2016ysw, DuttaBanik:2016jzv, Klasen:2016qux, Ghosh:2017fmr, Ahmed:2017dbb, Bhattacharya:2017fid, Ahmed:2017dbb, Borah:2017hgt, Bhattacharya:2018cqu, Bhattacharya:2018cgx, Aoki:2018gjf, DuttaBanik:2018emv, Barman:2018esi, YaserAyazi:2018lrv, Poulin:2018kap, Chakraborti:2018lso, Chakraborti:2018aae, Bernal:2018aon, Elahi:2019jeo, Borah:2019epq}. Such multi-component DM scenarios, even if both the DM candidates are of the same type, can have very interesting signatures at direct as well as indirect detection experiments, as studied in \cite{Profumo:2009tb, Fukuoka:2010kx, Cirelli:2010nh, Aoki:2012ub, Aoki:2013gzs, Modak:2013jya, Geng:2013nda, Gu:2013iy, Aoki:2014lha, Geng:2014zea, Geng:2014dea,  Biswas:2015sva, Borah:2015rla, Borah:2016ees,  Borah:2017xgm, Herrero-Garcia:2017vrl, Herrero-Garcia:2018mky}. Since direct and indirect detection (considering annihilations only, for stable DM) rates of DM are directly proportional 
to the DM density and DM density squared respectively and thereby producing tight constraints on single DM models, 
multi-component DM models can remain safe from being ruled out by such direct and indirect search constraints if the relative densities of different DM components are within appropriate limits. On top of that, such multi-component DM often comes with additional features like giving rise to interesting indirect detection signatures like monochromatic X-ray or gamma ray lines, as explored in several works, see for example \cite{Biswas:2013nn, Borah:2017hgt, Borah:2015rla, Borah:2016ees, Borah:2017xgm, DuttaBanik:2016jzv} and references therein.

Multi-component DM scenarios may also be connected with other sector of particle physics. One such immediate possibility 
evolves through a probable connection with neutrino physics, particularly with the origin of neutrino mass and mixing. 
Results of several experiments in last two decades like T2K \cite{Abe:2011sj}, Double Chooz \cite{Abe:2011fz}, Daya Bay 
\cite{An:2012eh}, RENO \cite{Ahn:2012nd} and MINOS \cite{Adamson:2013ue} have confirmed the existence of non-zero 
but tiny neutrino mass and large (compared to quark mixing) leptonic mixing \cite{Fukuda:2001nk, Ahmad:2002jz, Ahmad:2002ka, Abe:2008aa, Abe:2011sj, Abe:2011fz, An:2012eh, Ahn:2012nd, Adamson:2013ue, Olive:2016xmw}. Similar to the case 
of DM, these experimental observations provide clear indication for BSM physics as neutrino mass can not be explained 
within SM framework. Several BSM models attempt to explain tiny neutrino mass by incorporating additional fields. Apart 
from the conventional type I seesaw \cite{Minkowski:1977sc, GellMann:1980vs, Mohapatra:1979ia, Schechter:1980gr}, 
there exist other variants of seesaw mechanisms also, namely, type II seesaw \cite{Mohapatra:1980yp, Lazarides:1980nt, Wetterich:1981bx, Schechter:1981cv, Brahmachari:1997cq}, type III seesaw \cite{Foot:1988aq} and so on. 

It is particularly interesting to think of possible connection between the origin of neutrino mass and dark matter\cite{Bhattacharya:2016lts,Bhattacharya:2016rqj}, and perhaps such 
a connection is most straightforward in scotogenic scenarios, originally proposed by Ma \cite{Ma:2006km}. In scotogenic type of model, the $\mathbb{Z}_2$ odd particles take part in radiative generation of light neutrino masses, while the lightest $\mathbb{Z}_2$ odd particle plays the role of DM. The salient feature of this framework is the common origin of light neutrino mass and DM where one can constrain the model from observables in both the sectors, and hence enhancing the predictive power of the model. While there are several possible implementation of this scotogenic framework (for a review, please refer to \cite{Cai:2017jrq}) in single component DM scenarios, there have not been much studies on the role of multi-component DM on the origin of neutrino mass. For a recent work on the role of two component DM on radiative origin of neutrino mass, one may refer to \cite{Aoki:2017eqn} \footnote{In another recent work \cite{Bernal:2018aon}, two component fermion DM was proposed as a new anomaly free gauged $B-L$ model.}. We adopt a similar setup here while sticking to the most minimal scenario in order to accommodate two component scalar doublet DM with correct total relic abundance, satisfying the neutrino oscillation data and other relevant constraints from dark matter direct, indirect detections and lepton flavour violation (LFV).

We consider a $\mathbb{Z}_2 \times \mathbb{Z}^{\prime}_2$ extension of the Standard Model such that each $\mathbb{Z}_2$ sector consists of a singlet neutral fermion and a scalar field doublet under $SU(2)_L$. While the minimal scotogenic model consists of three singlet neutral fermion and one scalar doublet, our scenario contains two singlet fermions and two scalar doublets, keeping the particle content as minimal as the original model. Although there exists the possibility of singlet fermion DM also in either or both the $\mathbb{Z}_2$ sectors, we stick to scalar doublet DM. The reason behind this choice is three fold: (a) the gauge interactions of DM due to which the correct thermal abundance can be obtained easily without requiring large dimensionless couplings to enhance annihilation cross section\footnote{Fermion singlet DM in such models typically require large Yukawa couplings to satisfy correct relic and often run into the problems of vacuum stability \cite{Lindner:2016kqk}.}, (b) the gauge interactions of scalar doublet DM enhance its detection prospects at direct, indirect search experiments, (c) there exists an intermediate region for single component scalar doublet DM where relic abundance criteria can not be satisfied which makes it worth studying if two component scalar doublet DM can fill the void. Single component inert doublet DM (IDM) and its extensions have been studied by several authors in 
varieties of contexts \cite{Borah:2012pu, Dasgupta:2014hha, Cirelli:2005uq, Barbieri:2006dq, Ma:2006fn, LopezHonorez:2006gr,  Hambye:2009pw, Dolle:2009fn, Honorez:2010re, LopezHonorez:2010tb, Gustafsson:2012aj, Goudelis:2013uca, Arhrib:2013ela, Plascencia:2015xwa, Diaz:2015pyv,Ahriche:2017iar, Borah:2017dfn, Borah:2018uci, Borah:2018rca, Barman:2018jhz}.
We find that the total relic abundance of two component scalar doublet DM can be satisfied in the intermediate region while being consistent with neutrino oscillation data. The model also predicts the lightest neutrino mass to be zero. While the DM candidates satisfy the constraints from direct and indirect detection experiments, there lies the tantalising possibility to probe these scenarios at future searches in these frontiers and also in rare decay experiments like $\mu \rightarrow e \gamma$.

This paper is organised as follows. In section \ref{sec1}, we discuss the model and the particle spectrum. In section \ref{sec2} we discuss the details of two component dark matter pointing out the different annihilation channels contributing to the individual and total relic abundance, constraints from direct, indirect search followed by discussions on the constraints from neutrino oscillation data in section \ref{sec3}. We then briefly comment on lepton flavour violation in section \ref{sec31}. We discuss our results in section \ref{sec4} and finally conclude in section \ref{sec5}.

\section{The Model}
\label{sec1}
We have extended the particle content of the Standard Model by introducing two $SU(2)_L$ scalar doublets $\eta_1$ and 
$\eta_2$ and two right handed (RH) neutrinos, $N_{1,2}$. Furthermore, we include additional discrete symmetries, $\mathbb 
{Z}_2 \times \mathbb{Z}^{\prime}_2$ under which all SM fields transform trivially. The charge assignments of these additional 
fields under SM gauge symmetry as well as additional  global discrete symmetries are  indicated in Table \ref{qno}. The two 
neutral CP even scalars out of these extra doublets form the multi-component dark matter framework while the presence of RH neutrinos are instrumental in realising the light neutrino mass similar to that of scotogenic model \cite{Ma:2006km}. The additional discrete symmetries not only explain the stability of individual DM components, but also prevent Dirac neutrino mass at tree level by 
forbidding the couplings involving lepton doublets ($L$), singlet neutral fermions and the SM Higgs $H$.
The framework therefore serves as the minimal set-up in getting multi (two) component DM model which can accommodate 
light neutrino mass.  

The Yukawa Lagrangian involving interactions between the new fields of our model and SM fields can be written as
\bea
-\mathcal{L}^{\rm new} = Y_{\alpha 1} \bar{L}_{\alpha} \tilde{\eta}_1 N_1 + Y_{\alpha 2} \bar{L}_{\alpha} \tilde{\eta}_2 N_2 +\frac{1}{2} M_1 \bar{N_1^c} N_1 +\frac{1}{2} M_2 \bar{N_2^c} N_2 + h.c,
\label{lag} 
\eea
where $\alpha, \beta =  e, \mu, \tau$ stand for different generations of SM leptons. Note that as each RH neutrinos 
are odd under two different $\mathbb{Z}_2$ sectors, the corresponding RH neutrino mass matrix remains diagonal.

\begin{table}[t]
\centering
\begin{tabular}{|c c c c|}
\hline
Field   & ~~~~$SU(3)_c \times SU(2)_L \times U(1)_Y$  & ~~~~$\mathbb{Z}_2$ & ~~~~${\mathbb{Z}_2}^{\prime}$\\ \hline \hline
$\eta_1$ & (1, 2,$~\frac{1}{2}$) & - & +\\ \hline
$\eta_2$ & (1, 2,$~\frac{1}{2}$) & + & -\\ \hline
$N_1$ & (1, 1,$0$) & - & +\\ \hline
$N_2$ & (1, 1,$0$) & + & -\\ \hline
\end{tabular}
\caption{New particle content of the model and their charge assignments.}
\label{qno}
\end{table}

The most general renormalisable scalar potential of our model, $V(H, \eta_1, \eta_2)$, consistent with $SU(2)_L \times U(1)_Y \times \mathbb{Z}_2\times \mathbb{Z}_2^{\prime}$ consists of (i) $V_H$: sole contribution of the SM Higgs , (ii) $V_{\eta_1}$: $\eta_1$ contribution, (iii) $V_{\eta_2}$: $\eta_2$ contribution and (iv) $V_{\rm{int}}$: interactions among $H, \eta_1, \eta_2$. This can be written as follows.
\bea
V(H,\eta_1,\eta_2) =  V_{H} + V_{\eta_1} + V_{\eta_2} + V_{\rm {int}},
\eea
where 
\besub
\bea
V_{H} &=& -\mu_H^2 H^{\dagger} H + \l_H (H^{\dagger} H)^2,  \\
V_{\eta_1} &=& \mu_1^2 \eta_1^{\dagger} \eta_1
+ \l_{\eta_1} (\eta_1^{\dagger} \eta_1)^2,  \\
V_{\eta_2} &=& 
\mu_2^2 \eta_2^{\dagger} \eta_2+ \l_{\eta_2} (\eta_2^{\dagger} \eta_2)^2, 
\eea
\eesub
and 
\bea
V_{\rm{int}}&=& 
 \l_{3} (H^{\dagger} H) (\eta_1^{\dagger} \eta_1)
 + \l_{4} (H^{\dagger} \eta_1) (\eta_1^{\dagger} H)
 + \frac{\l_{5}}{2} \Big[(H^{\dagger} \eta_1)^2 + 
 (\eta_1^{\dagger} H)^2 \Big] \nonumber \\
&&
 + \tilde{\l_{3}} (H^{\dagger} H) (\eta_2^{\dagger} \eta_2)
 + \tilde{\l_{4}} (H^{\dagger} \eta_2) (\eta_2^{\dagger} H)
 + \frac{\tilde{\l_{5}}}{2} \Big[(H^{\dagger} \eta_2)^2 + 
 (\eta_2^{\dagger} H)^2 \Big] \nonumber \\
&&
 + \l_{3}^{\prime} (\eta_1^{\dagger} \eta_1) (\eta_2^{\dagger} \eta_2)
 + \l_{4}^{\prime} (\eta_1^{\dagger} \eta_2) (\eta_2^{\dagger} \eta_1)
 + \frac{\l_{5}^{\prime}}{2} \Big[(\eta_1^{\dagger} \eta_2)^2 + 
 (\eta_2^{\dagger} \eta_1)^2 \Big]. \nonumber \\ 
 \eea

In order to keep $\mathbb{Z}_2\times \mathbb{Z}_2^{\prime}$ unbroken so as to guarantee the stability of DM components, 
the neutral components of $\eta_1$ and $\eta_2$ are chosen not to acquire any non-zero vacuum expectation value (VEV) and hence they can be identified as two inert Higgs doublets (IHD). These IHDs can be parametrised as
\bea
\eta_1 &=&
  \begin{pmatrix}
    \eta_1^+\\
 \frac{1}{\sqrt 2}(H_1 + i A_1)
  \end{pmatrix},
\eta_2 =
  \begin{pmatrix}
    \eta_2^+\\
 \frac{1}{\sqrt 2}(H_2 + i A_2)
  \end{pmatrix}.
 \eea
On the other hand, the neutral component of the SM Higgs field acquires a non-zero VEV (denoted by $v$) which is responsible for electroweak symmetry breaking (EWSB). We parametrise the Higgs field $H$ as
\bea
H &=&
  \begin{pmatrix}
    0 \\
 \frac{1}{\sqrt 2}(v+h)
  \end{pmatrix}.
\eea As follows from the scalar field Lagrangian, after EWSB, the physical scalars have the following masses
\besub
\bea
m_{h}^2 &=& \frac{1}{2}\l_{H}v^2,  \\
m_{H_{1}}^2 &=& 
\mu_1^2 + \frac{1}{2}(\l_{3} + \l_{4} + \l_{5})v^2, \\
m_{A_{1}}^2 &=& 
\mu_1^2 + \frac{1}{2}(\l_{3} + \l_{4} - \l_{5})v^2, \\
m_{\eta_{1}^+}^2 &=& 
\mu_1^2 + \frac{1}{2} \l_{3} v^2, \\
m_{H_{2}}^2 &=& 
\mu_2^2 + \frac{1}{2}(\tilde{\l_{3}} +\tilde{\l_{4}} + \tilde{\l_{5}})v^2,  \\
m_{A_{2}}^2 &=& 
\mu_2^2 + \frac{1}{2}(\tilde{\l_{3}} + \tilde{\l_{4}} - \tilde{\l_{5}})v^2, \\
m_{\eta_{2}^+}^2 &=& 
\mu_2^2 + \frac{1}{2} \tilde{\l_{3}} v^2  .
\eea
\label{eq5}
\eesub
The scalar potential should be bounded from below in order to make the electroweak vacuum stable. This poses some 
constraints on the scalar couplings of the model. In addition to this, all the relevant couplings should also maintain the 
perturbativity. These bounds together with the unitarity limit have been studied for the three Higgs doublets and can 
be found in \cite{Chakrabarty:2015kmt}, from which we obtain the limits in our case.

We are effectively left with two-component inert DM  scenario where the presence of light neutrinos are also taken cared of. 
As specified before, we consider the CP even neutral components $H_1, H_2$ as two DM candidates 
without any loss of generality. Analogous to the single IHD scenario, we define $\l_{L1} \equiv \frac{\l_{3} + \l_{4} + \l_{5}}{2}$ and $\l_{L2} \equiv \frac{\tilde{\l_{3}} + \tilde{\l_{4}} + \tilde{\l_{5}}}{2}$, which denote the individual Higgs portal couplings of two DM candidates respectively. For our analysis purpose, we first implement this model in \texttt{LanHEP} \cite{Semenov:2014rea} choosing the independent parameters in the scalar sector as 
$$(m_{H_1}, m_{A_1},m_{\eta_{1}^+},m_{H_2}, m_{A_2},m_{\eta_{2}^+},\l_{L1},\l_{L2},\l_{\eta_1},\l_{\eta_2},\l_{3}^{\prime},\l_{4}^{\prime},\l_{5}^{\prime}).$$  For simplicity, we will consider the couplings of quartic interaction between 
two IHDs ($\eta_1$ and $\eta_2$) to be the same and identify it by $\l_{12}$, $i.e. \l_{3}^{\prime},\l_{4}^{\prime},\l_{5}^{\prime} = \l_{12}$. 
We express other couplings of the scalar sector in terms of these independent parameters as follows,
\besub
\bea
\mu_1^2 &=& m_{H_1}^2 - \l_{L1} v^2,   \\
\l_{3} &=& 2\l_{L1} + \frac{2(m_{\eta_{1}^+}^2-m_{H_{1}}^2 )}{v^2}, \\
\l_{4} &=& \frac{m_{H_{1}}^2 + m_{A_1}^2 - 2 m_{\eta_{1}^+}^2}{v^2}, \\
\l_{5} &=& \frac{(m^2_{H_1} - m^2_{A_1})}{v^2}, \\
\mu_2^2 &=& m_{H_2}^2 - \l_{L2} v^2,   \\
\tilde{\l_{3}} &=& 2\l_{L2} + \frac{2(m_{\eta_{2}^+}^2-m_{H_{2}}^2 )}{v^2}, \\
\tilde{\l_{4}} &=& \frac{m_{H_{2}}^2 + m_{A_2}^2 - 2 m_{\eta_{2}^+}^2}{v^2}, \\
\tilde{\l_{5}} &=& \frac{(m^2_{H_2} - m^2_{A_2})}{v^2}. 
\eea
\eesub

Note that the parameters $\l_{\eta_1}$ and $\l_{\eta_2}$ do not take part directly in the DM phenomenology.

\section{Dark Matter Phenomenology}
\label{sec2}

The set-up contains two dark matter components: $H_1$ and $H_2$. In order to find the final relic density, their annihilations 
and co-annihilations are to be considered. In addition, the role of neutrino Yukawa couplings are also important. Below we 
provide a systematic approach to calculate the relic density in our scenario. Constraints from dark matter search are 
also mentioned. 

\subsection{\textbf{Relic Density}}
For a single component WIMP type DM, the DM candidate with mass $m_{DM}$ is part of the thermal plasma in the early 
universe which eventually freezes out when the rate of annihilations fall below the rate of expansion of the universe. The 
final abundance can then be obtained by solving the Boltzmann equation for the DM number density. In fact, for DM 
annihilations dominated by $s$-wave processes only, the relic abundance can be approximated as ~\cite{Kolb:1990vq}
\begin{align}
\Omega_{\rm DM}h^2 \ = \ \left[1.07\times 10^9~{\rm GeV}^{-1}\right]\frac{ x_fg_*^{1/2}}{g_{*s}M_{\rm Pl}\langle \sigma v\rangle_f} \, ,
\label{eq:omega}
\end{align}
where $g_*$ and $g_{*s}$ are the effective relativistic degrees of freedom that contribute to the energy density and entropy density, respectively. $x_f$ is to be determined from the parameter $x = m_{DM}/T$ evaluated at the freeze-out temperature 
$T_f$ and is given by   
$x_f \equiv [\frac{m_{\rm DM}}{T}]_{T = T_f} \ = \ \ln \left(0.038\frac{g}{g_*^{1/2}}M_{\text{Pl}}m_{\rm DM}\langle \sigma v\rangle_f\right)$, 
with $g$ being the number of internal degrees of freedom of the DM. The thermally averaged annihilation cross section, 
given by~\cite{Gondolo:1990dk}
\begin{equation}
\langle \sigma v \rangle \ = \ \frac{1}{8m_{\rm DM}^4T K^2_2\left(\frac{m_{\rm DM}}{T}\right)} \int\limits^{\infty}_{4m_{\rm DM}^2}\sigma (s-4m_{\rm DM}^2)\sqrt{s}\: K_1\left(\frac{\sqrt{s}}{T}\right) ds \, ,
\label{eq:sigmav}
\end{equation}
is evaluated at $T_f$ and denoted by $\langle \sigma v \rangle_f$. The freeze-out temperature $T_f$ is derived from 
the equality condition of DM interaction rate $\Gamma = n_{\rm DM} \langle \sigma v \rangle$ with the rate of expansion of the universe $H(T) \simeq \sqrt{\frac{\pi^2 g_*}{90}}\frac{T^2}{M_{\rm Pl}}$. In the above expression of Eq.(\ref{eq:sigmav}), $K_i(x)$'s are the modified Bessel functions of order $i$. As is well known,  if the mass splitting within the DM multiplet is relatively small, there can be additional contributions from coannihilation channels \cite{Griest:1990kh}, whose importance in IDM has 
already been discussed in several earlier works.

In the present model, we have two dark matter candidates  $H_1$ and $H_2$. Since both the candidates contribute to 
the dark matter relic density obtained from Planck\cite{Aghanim:2018eyx} experiment, one must satisfy the following 
relation:
\bea
\Omega_{\rm DM}h^2 = 0.120\pm 0.001 = \Omega_{1}h^2 + \Omega_{2}h^2
\label{relic_total} ,
\eea
where $h$ denotes the reduced Hubble parameter and the relic density of the $H_1$ and $H_2$ is given by $\Omega_{1}h^2$ and $\Omega_{2}h^2$ respectively. Since there exists annihilation channels through which $H_1$ can go into $H_2$ (or vice versa, depending upon which one is heavier), the Boltzmann equations for the two DM candidates are coupled, in general. In such a case, there is no approximate formula which we can use for individual DM abundance like we had in case of single component DM discussed above. 

\begin{figure}[H]
\centering
\subfigure[]{
\includegraphics[scale=0.30]{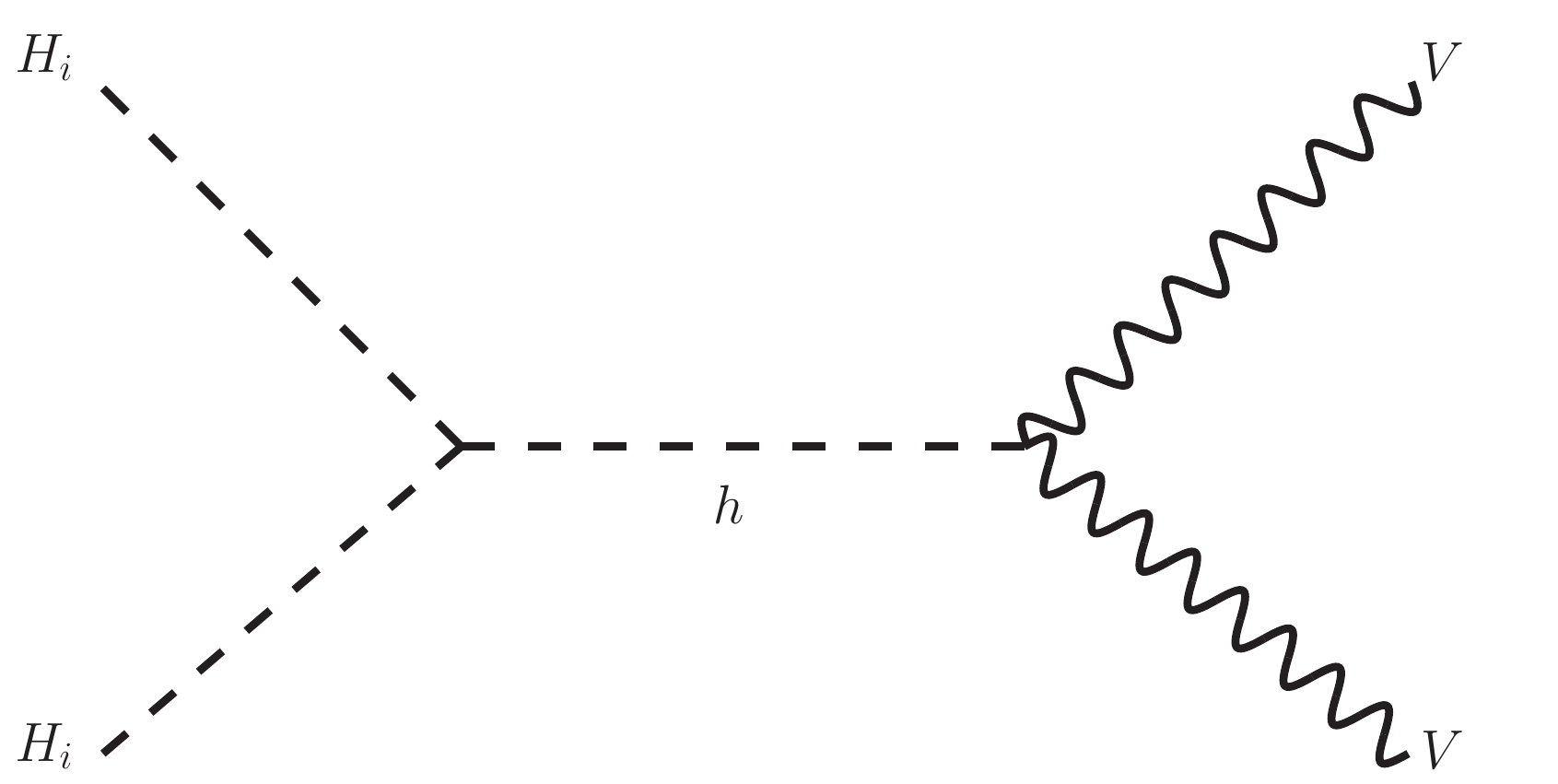}}
\subfigure[]{
\includegraphics[scale=0.30]{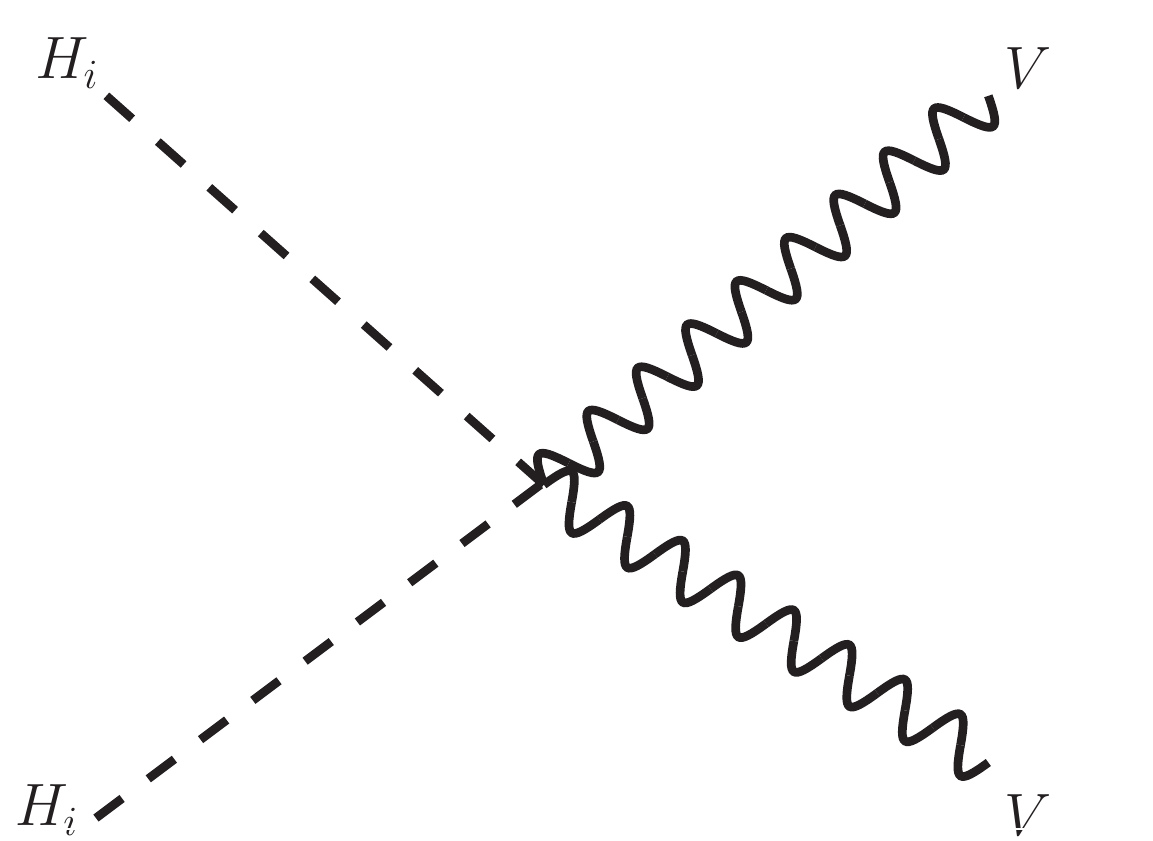}}
\subfigure[]{
\includegraphics[scale=0.30]{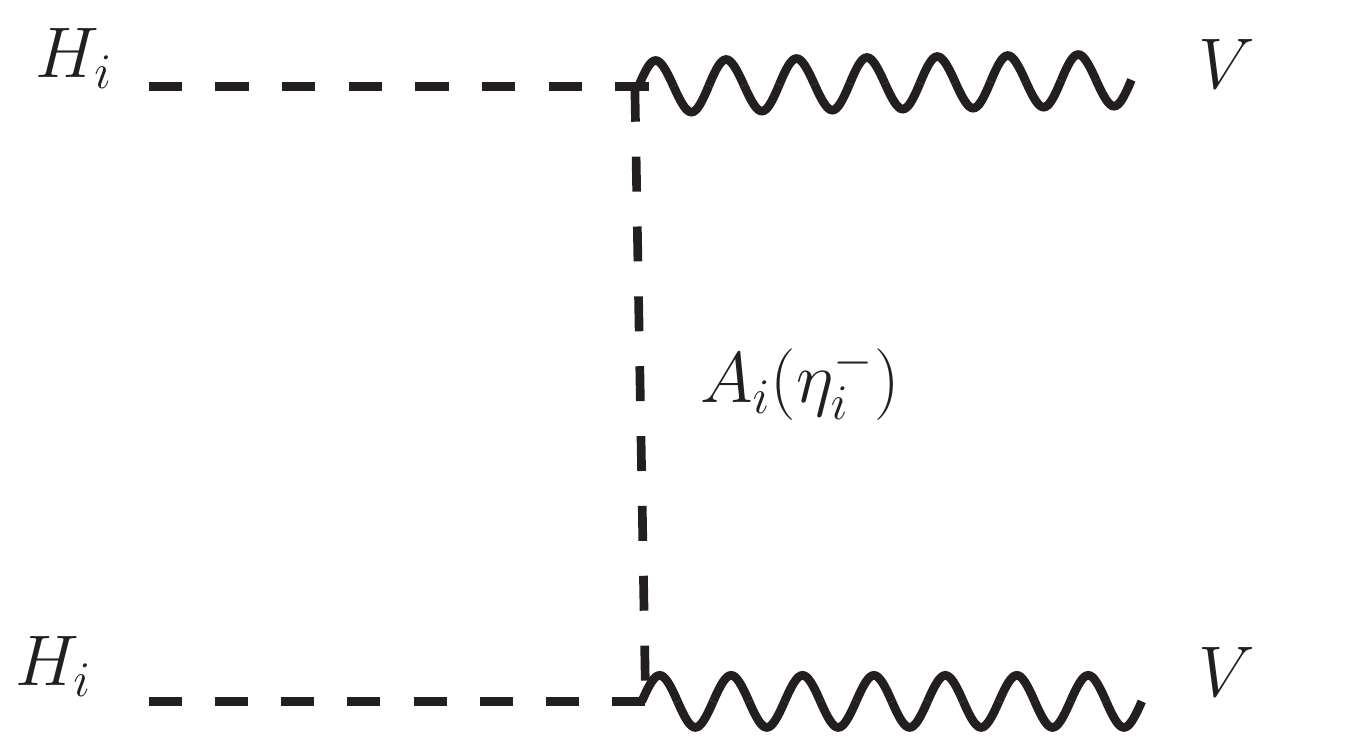}}
\subfigure[]{
\includegraphics[scale=0.30]{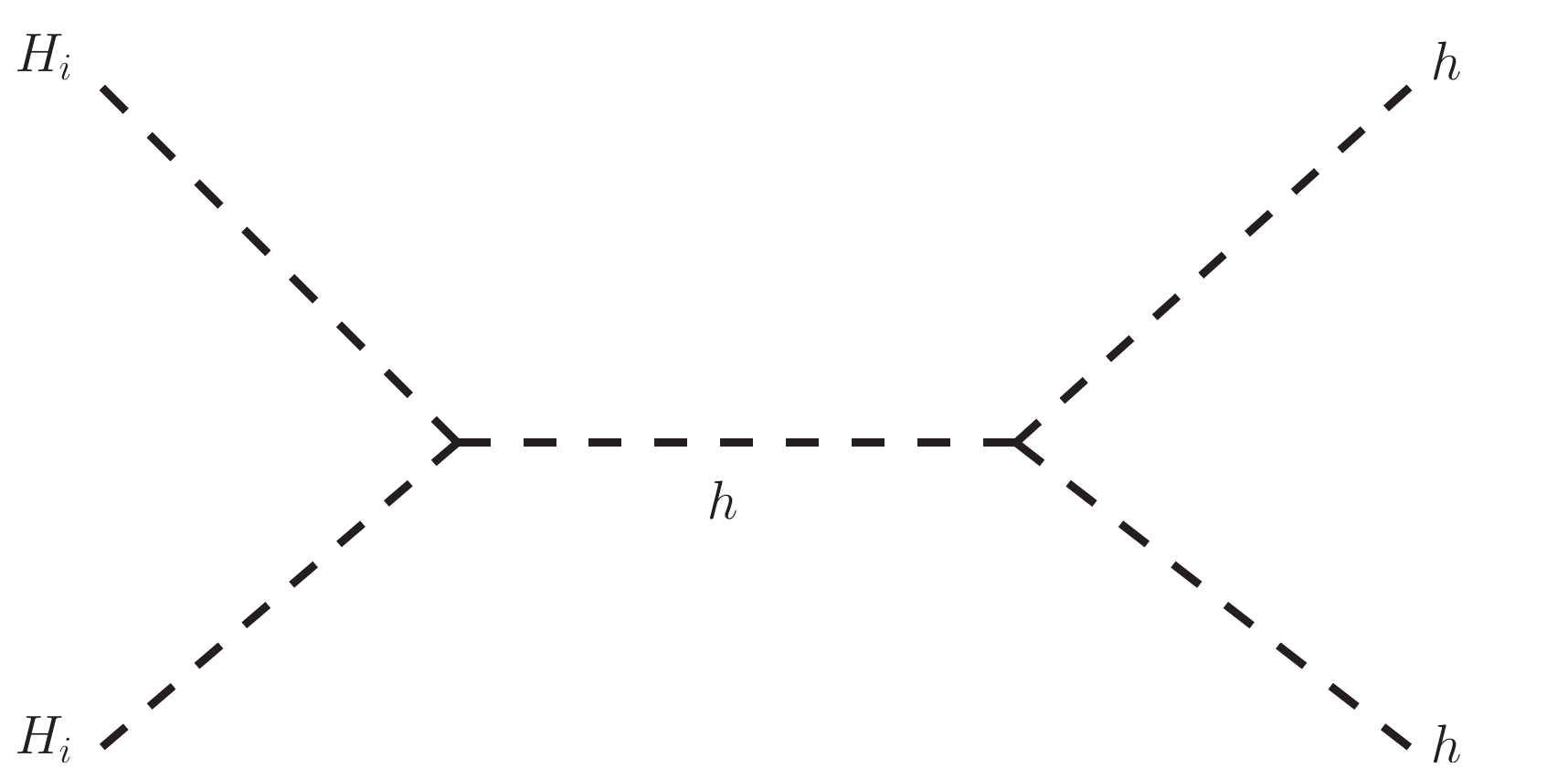}}
\subfigure[]{
\includegraphics[scale=0.30]{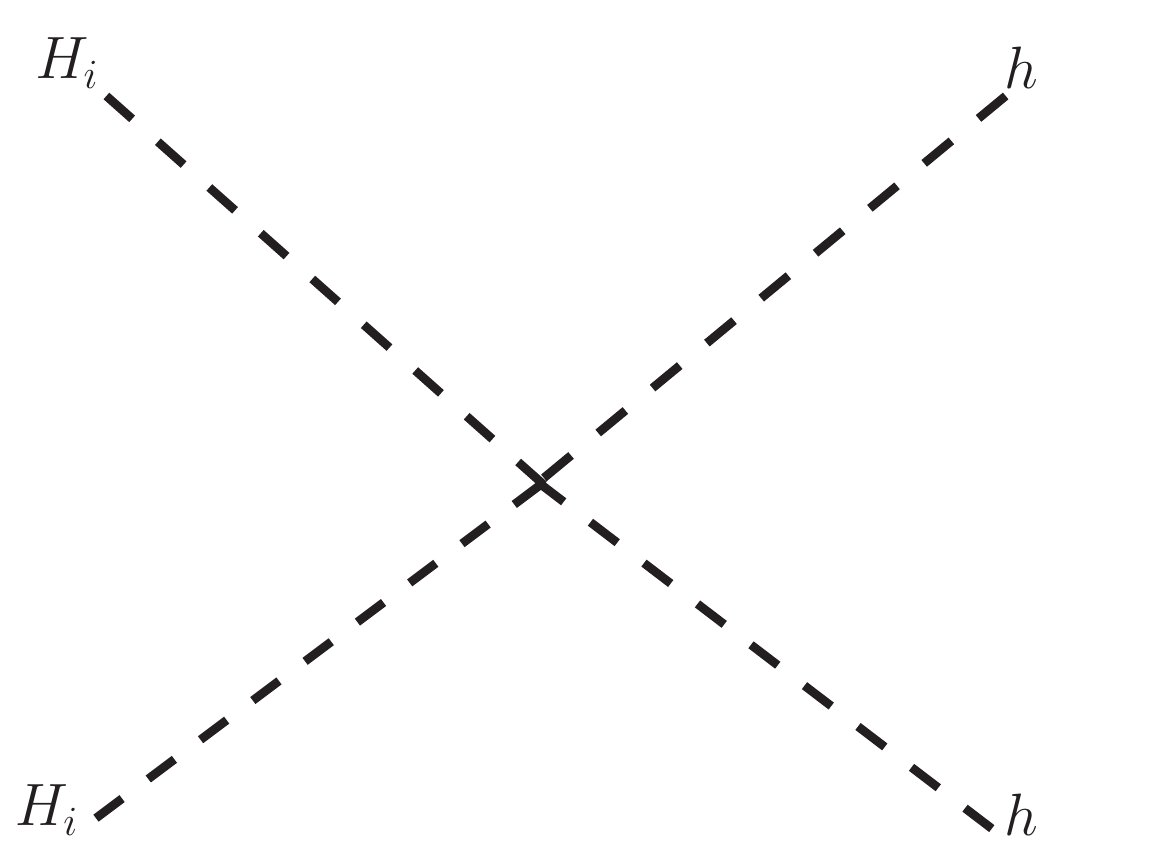}}
\subfigure[]{
\includegraphics[scale=0.30]{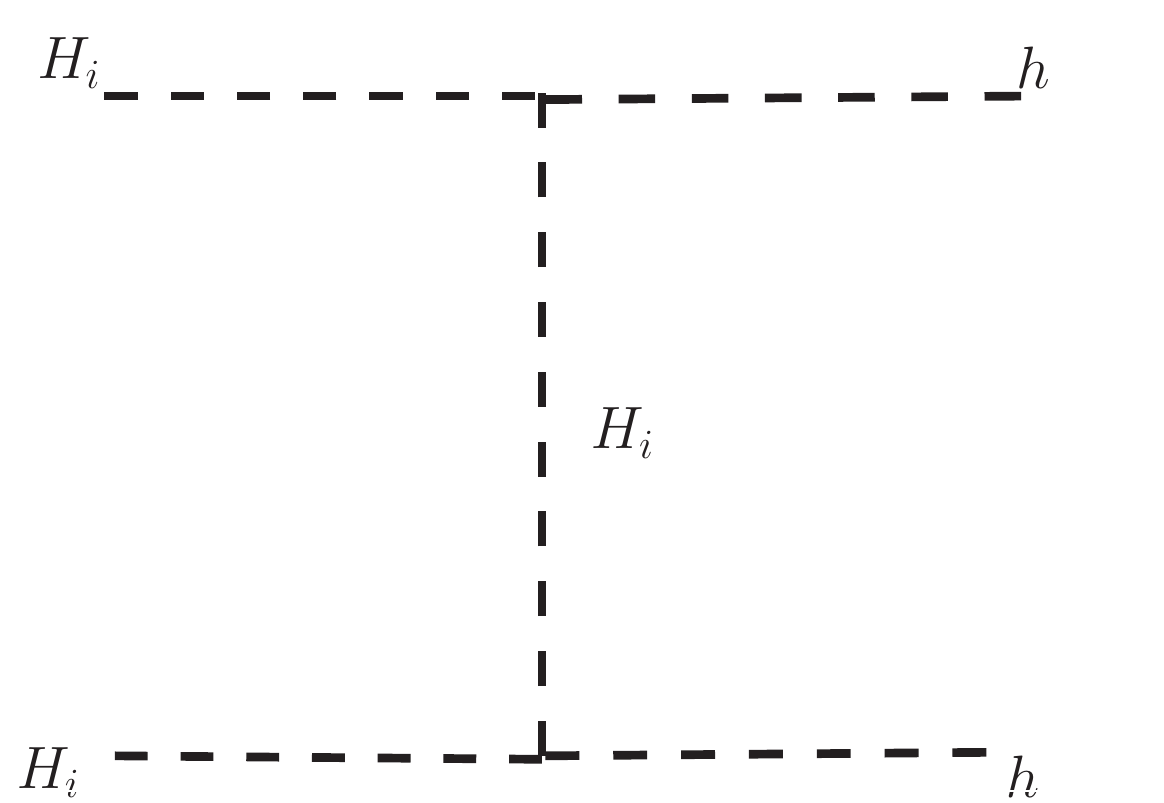}}
\subfigure[]{
\includegraphics[scale=0.30]{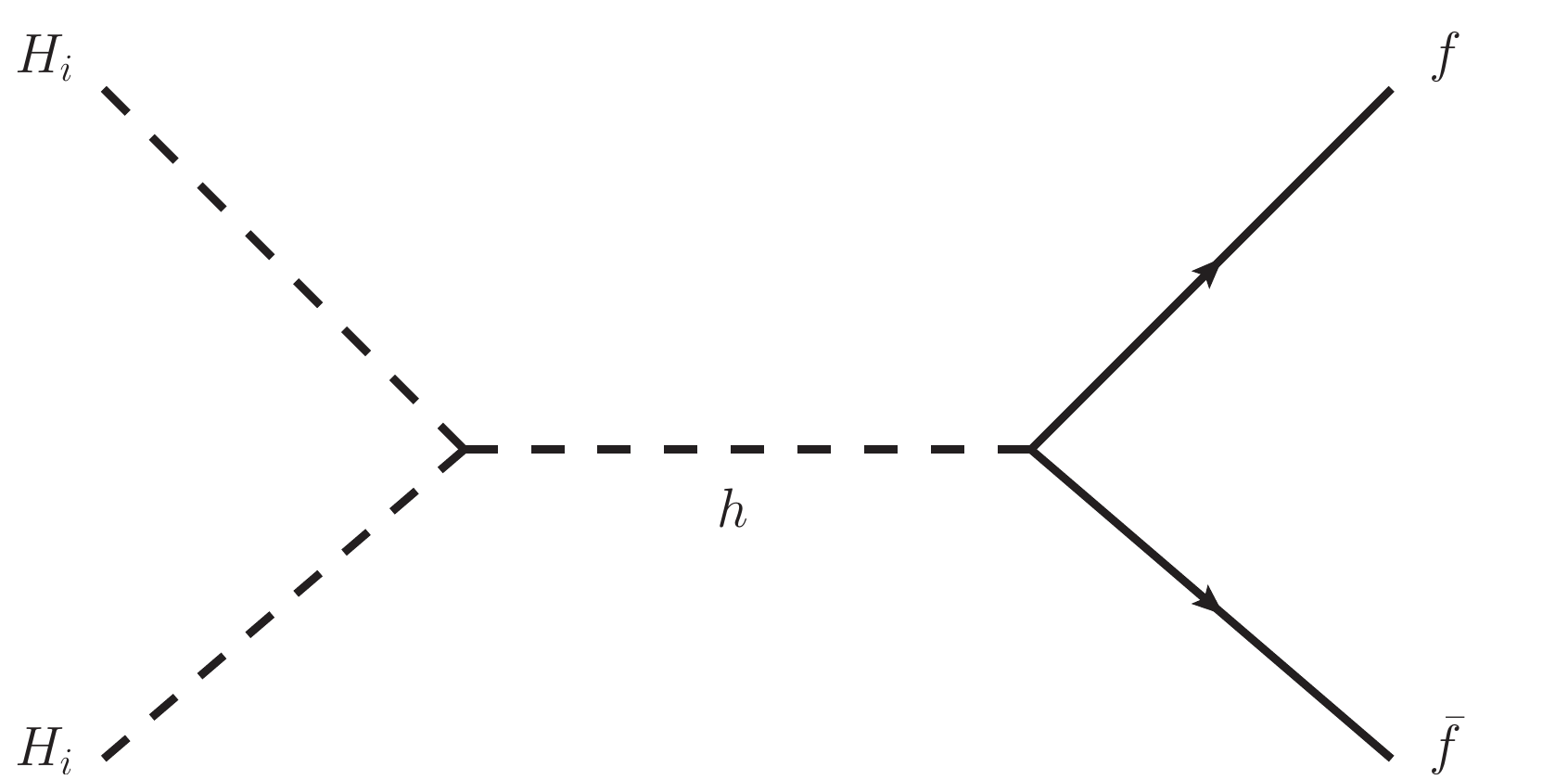}}
\caption{Annihilation channels}
\label{feyn1}
\end{figure}

\begin{figure}[H]
\centering
\subfigure[]{
\includegraphics[scale=0.30]{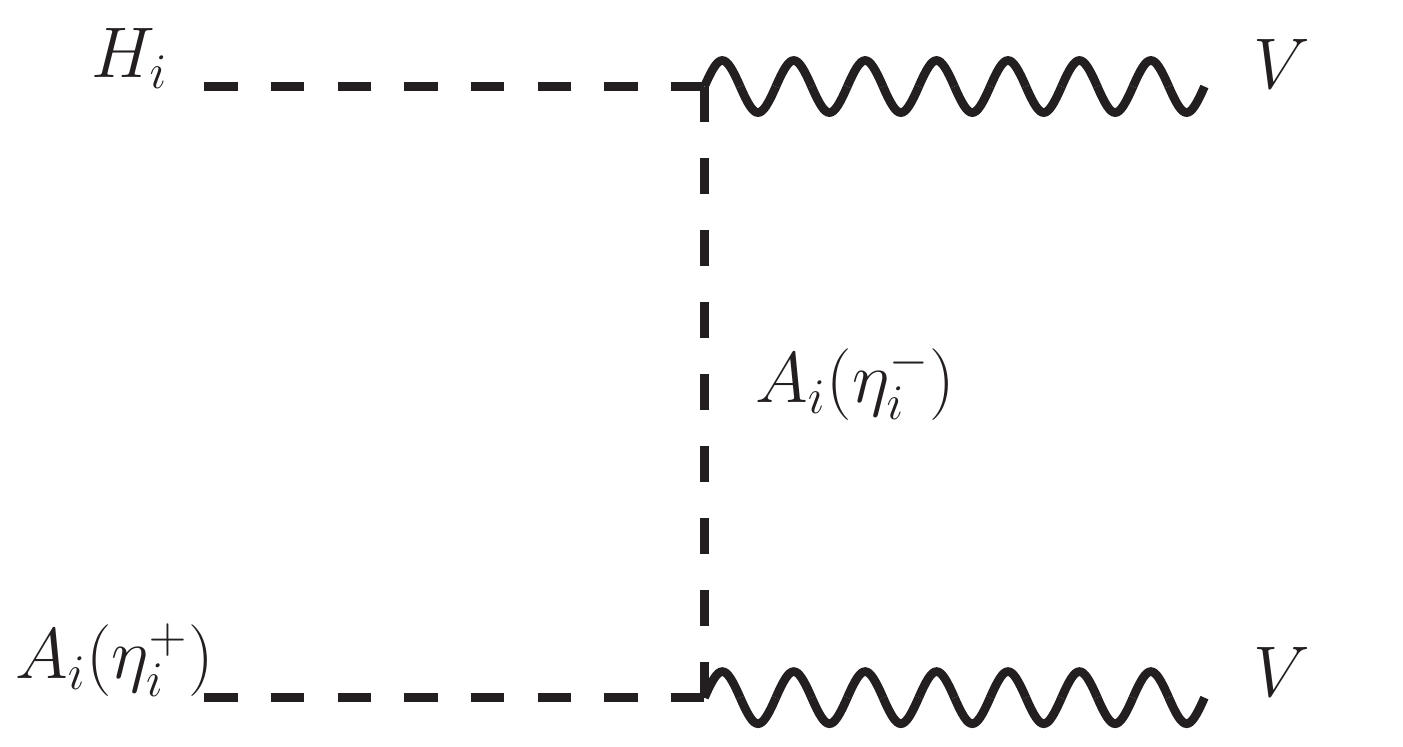}}
\subfigure[]{
\includegraphics[scale=0.30]{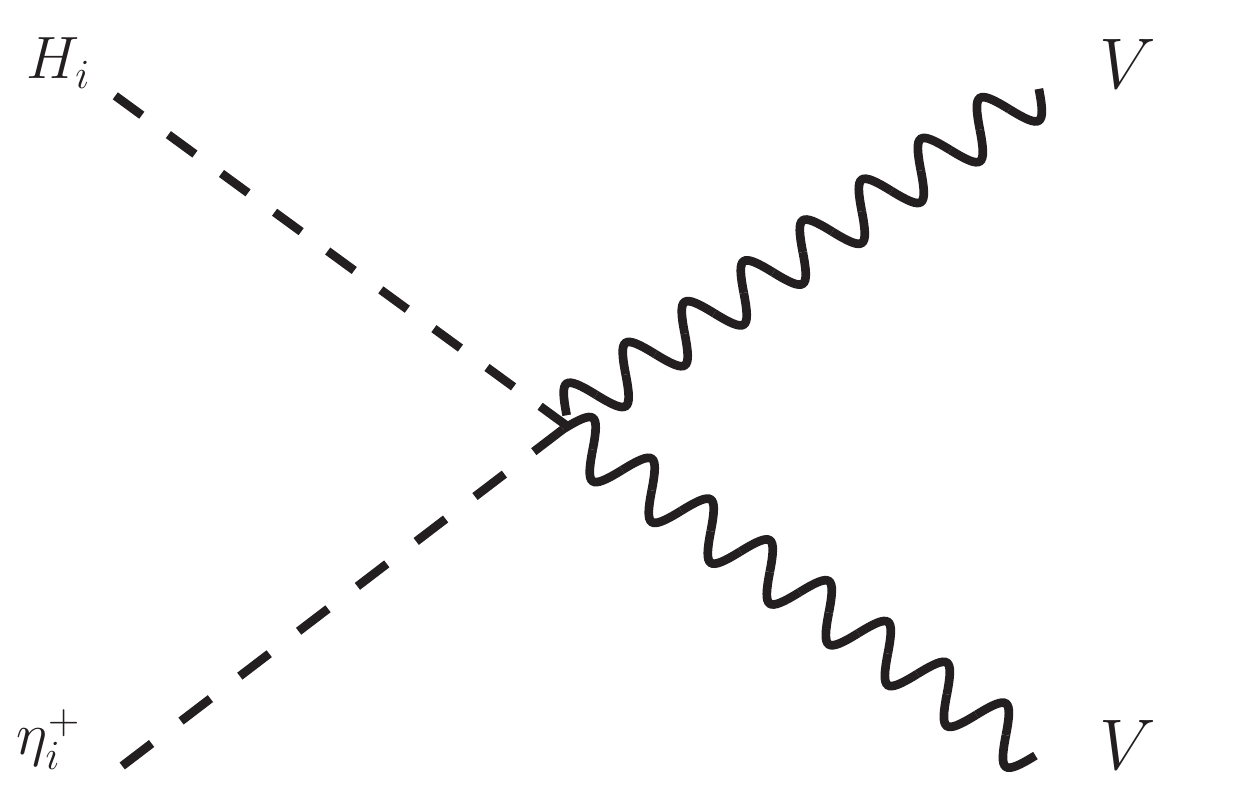}}
\subfigure[]{
\includegraphics[scale=0.30]{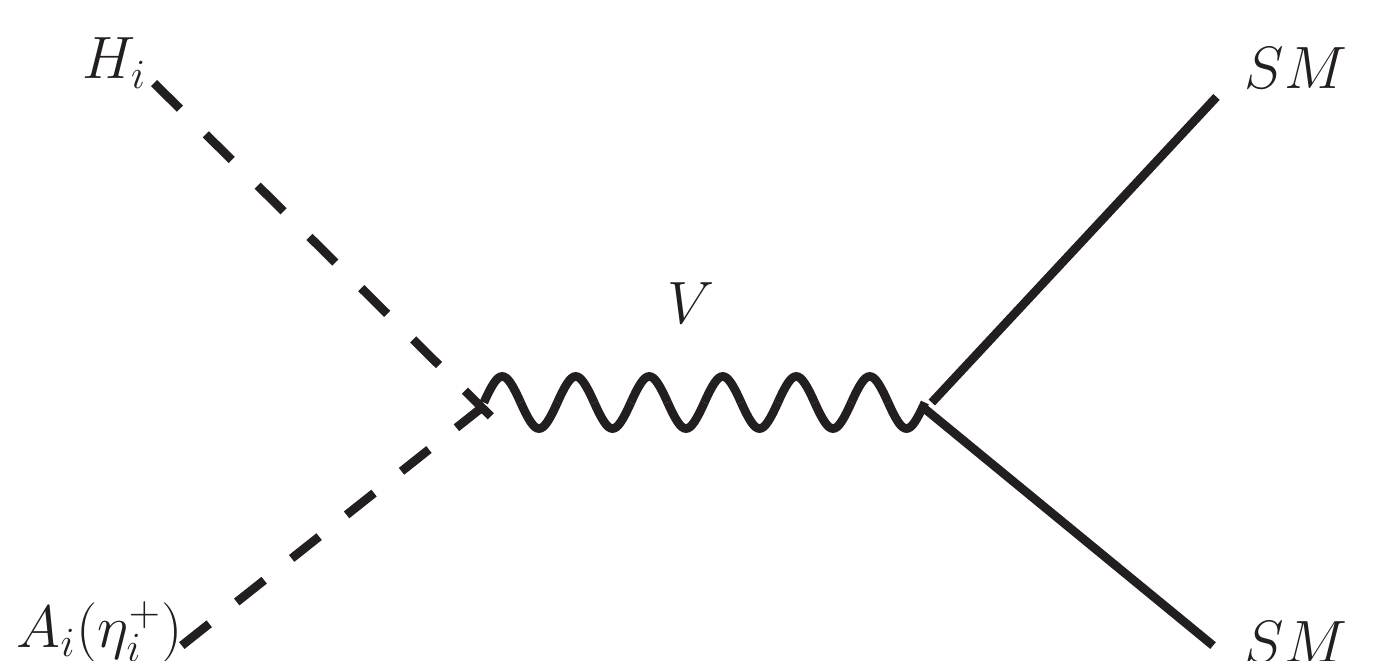}}
\caption{Coannihilation Channels}
\label{feyn01}
\end{figure}

\begin{figure}[H]
\centering
\subfigure[]{
\includegraphics[scale=0.30]{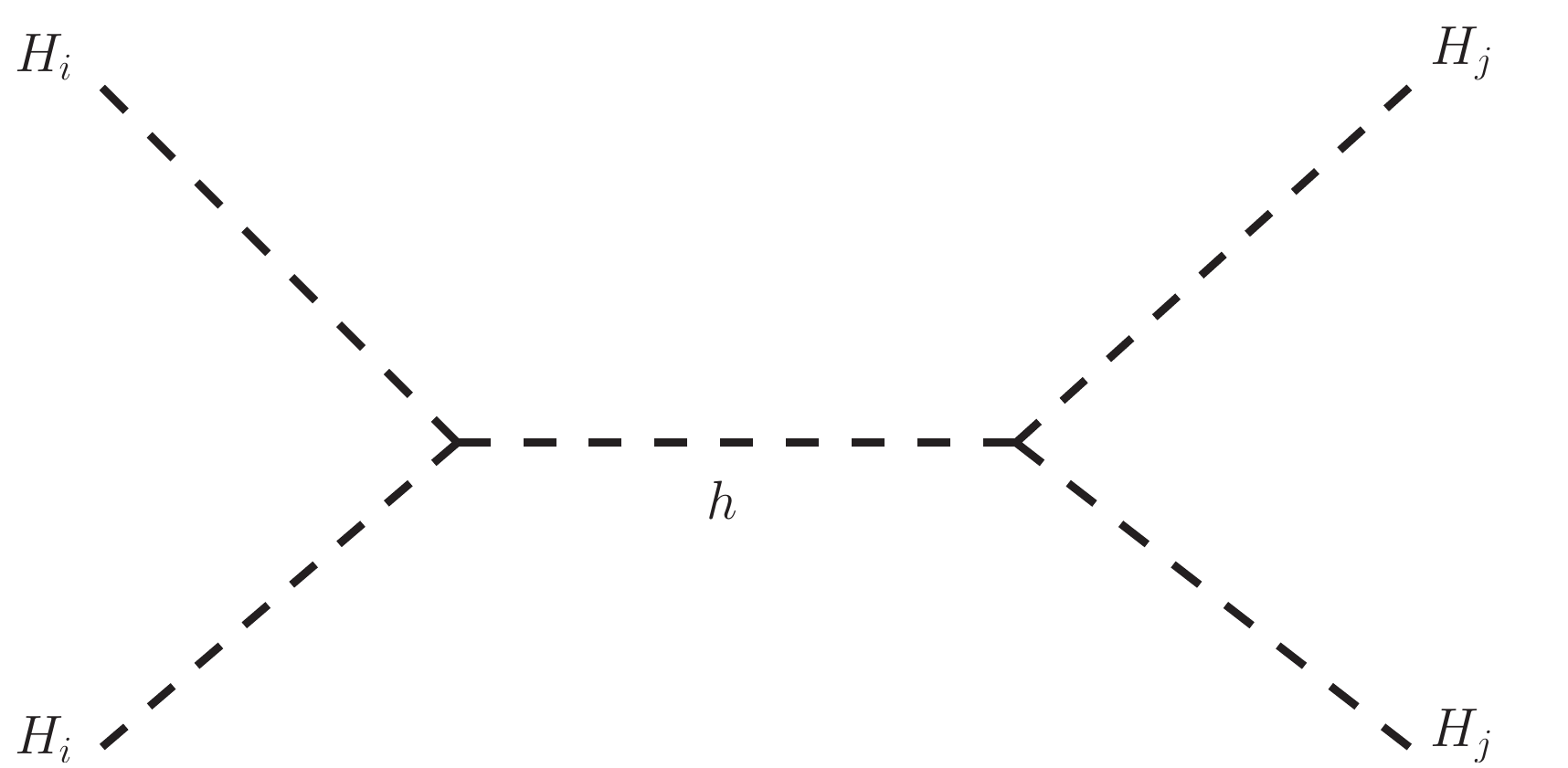}}
\subfigure[]{
\includegraphics[scale=0.30]{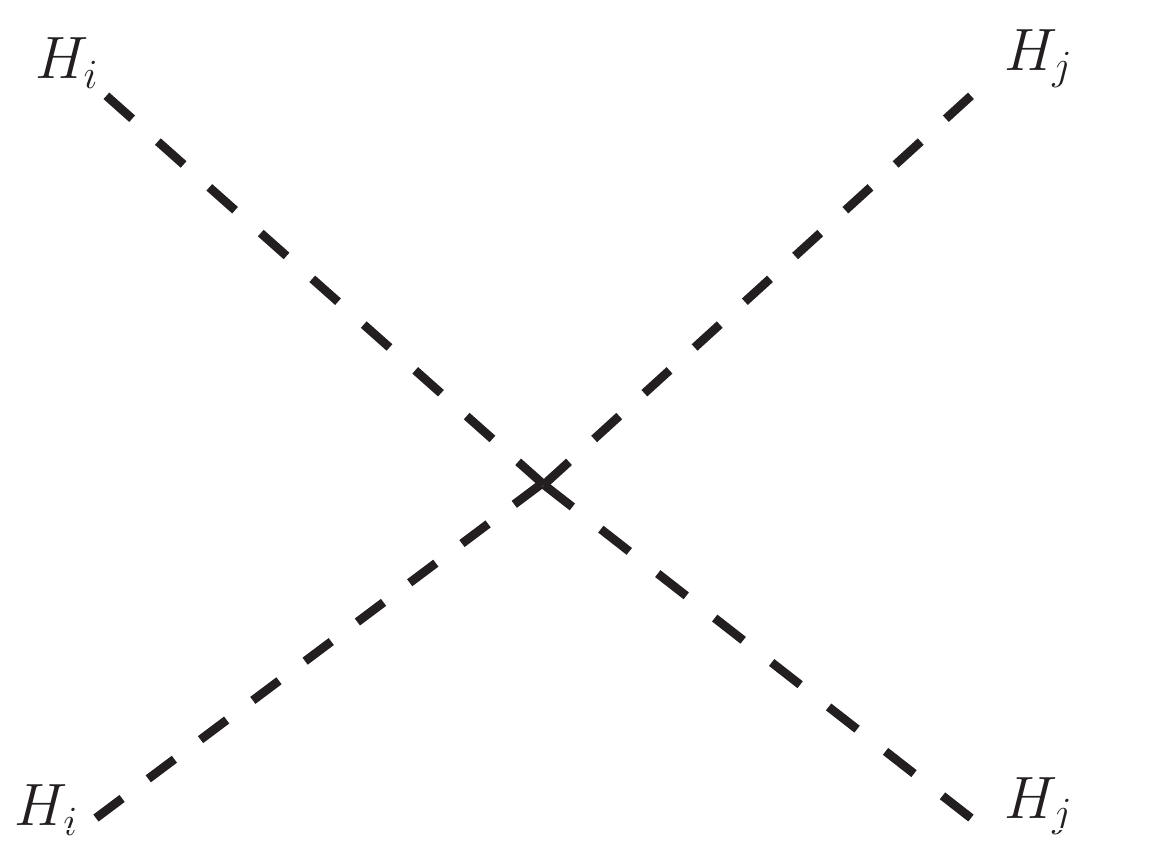}}
\caption{DM-DM conversion Channels}
\label{feyn2}
\end{figure}

\begin{figure}[H]
\centering
\subfigure[]{
\includegraphics[scale=0.30]{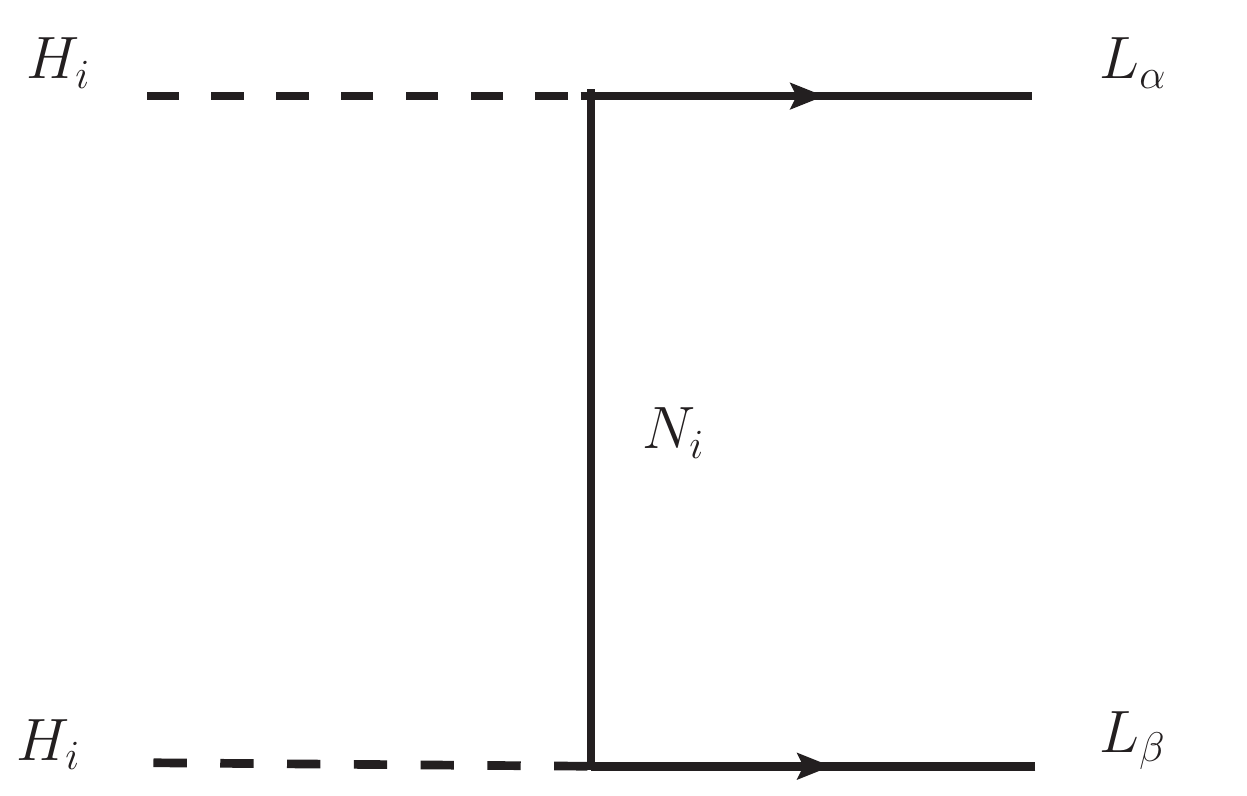}}
\subfigure[]{
\includegraphics[scale=0.30]{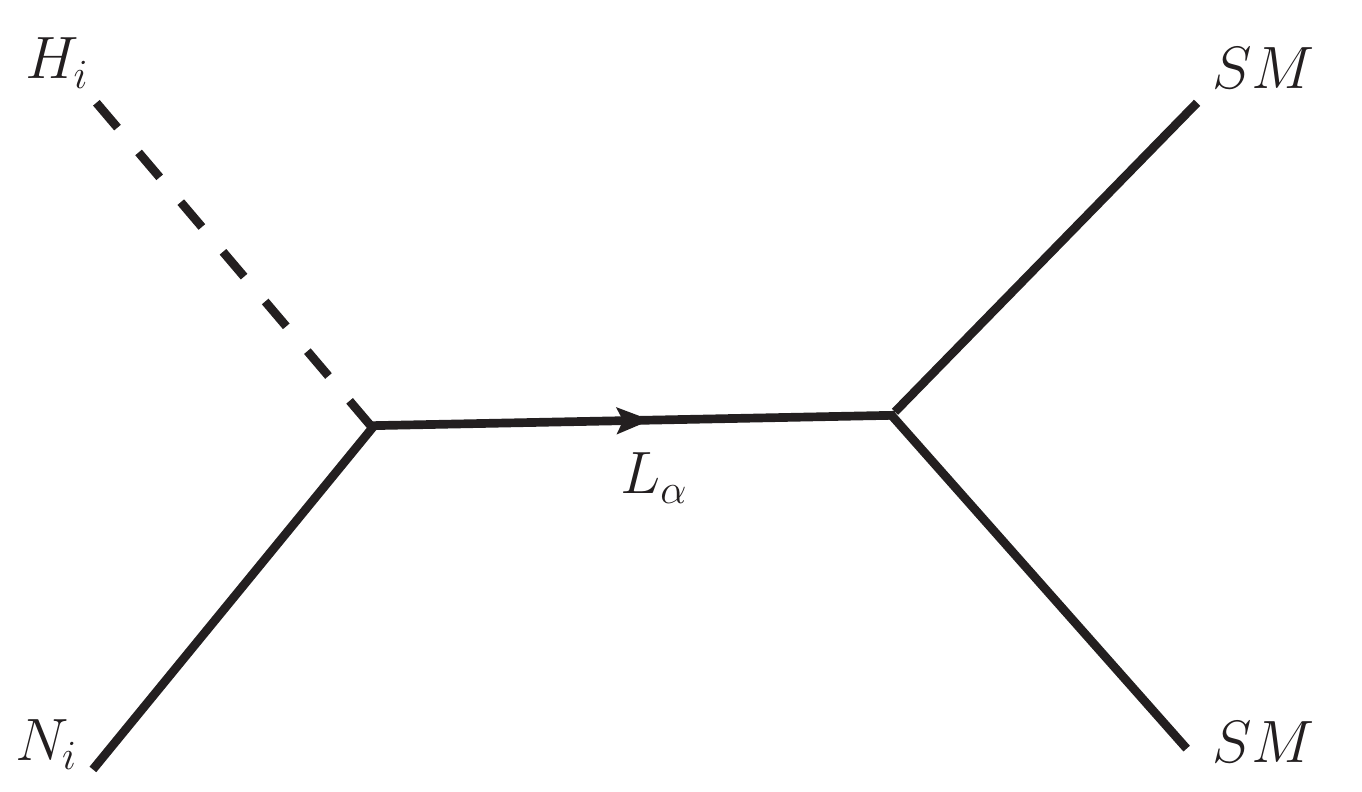}}
\subfigure[DM-DM conversion]{
\includegraphics[scale=0.30]{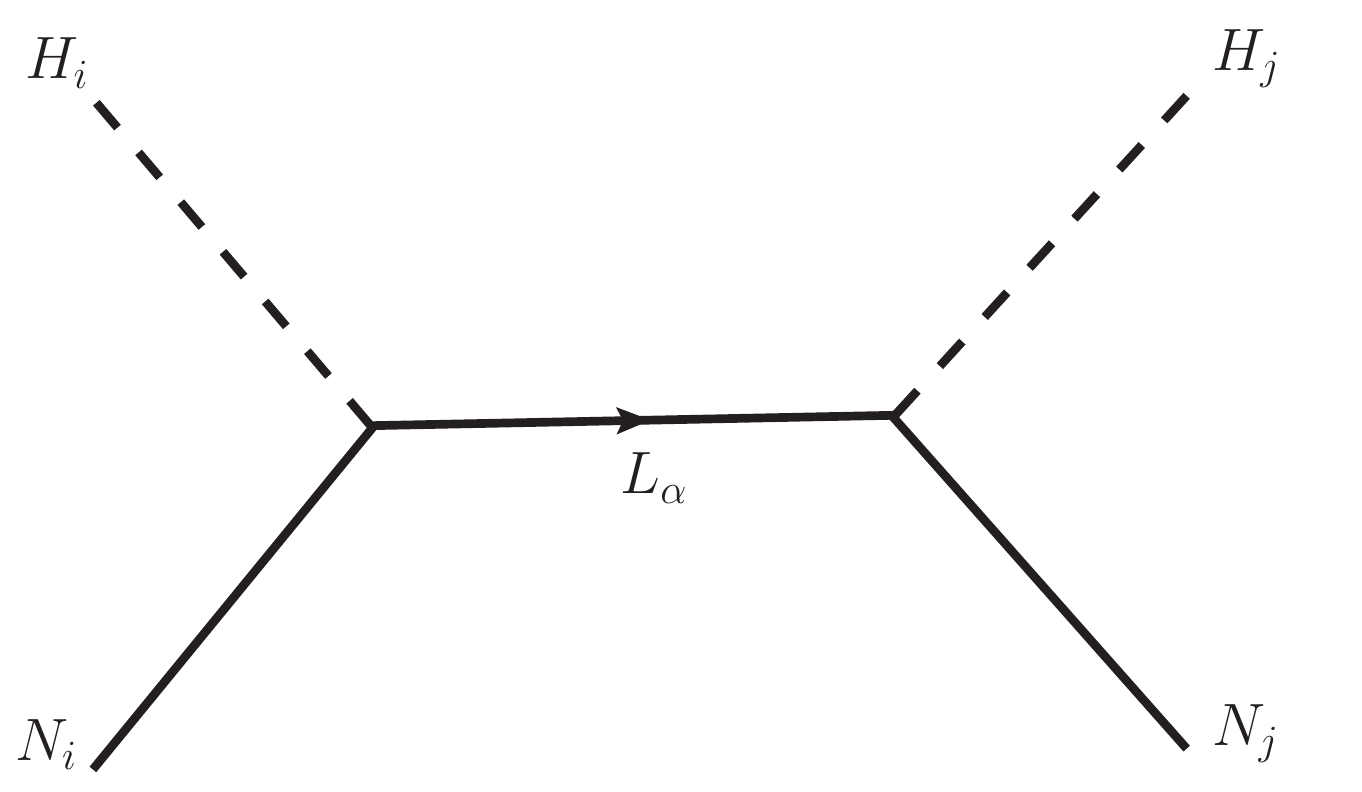}}
\subfigure[DM-DM conversion]{
\includegraphics[scale=0.30]{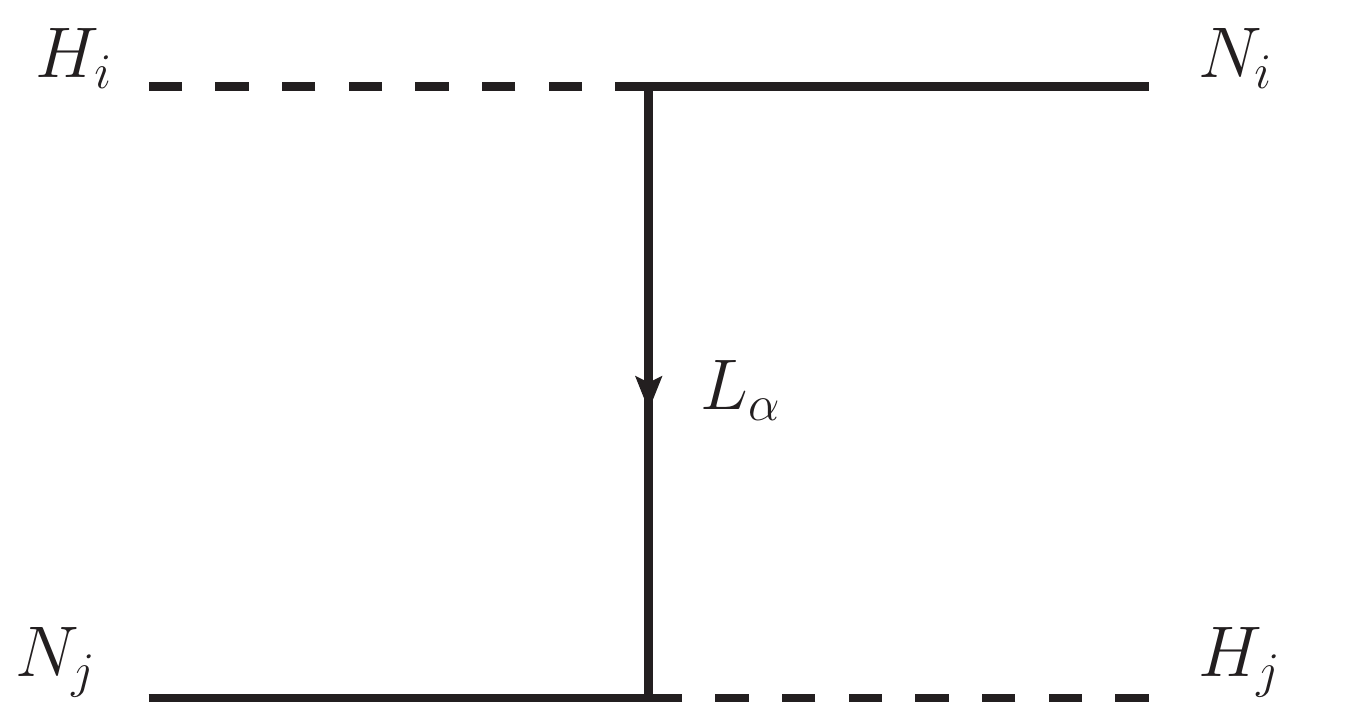}}
\caption{(Co)annihilation channels in presence of singlet neutral fermions}
\label{feyn1a}
\end{figure}

Before going to the details of coupled Boltzmann equations, we first identify and categorise different annihilation 
channels of $H_1, H_2$ as shown in Fig. \ref{feyn1}, co-annihilations channels involving individual DM candidates 
in  Fig. \ref{feyn01}, along with conversion between DM candidates $H_1, H_2$ (depending upon which one is 
heavier) in Fig. \ref{feyn2}.  Feynman diagrams including contribution from neutrino Yukawa interactions are included 
in Fig. \ref{feyn1a}. 

First we write down the coupled Boltzmann equations without considering the effect of neutrino Yukawa coupling ($i.e.$ excluding 
Fig. \ref{feyn1a}). Due to the involvement of two components of DM in our case, we modify the definition of parameter $x$ 
from before as $x = \mu/T$, where $\mu$ is the reduced mass defined through:$~\mu= \frac{m_{H_1}m_{H_2}}{m_{H_1}+m_{H_2}}$. Therefore the coupled Boltzmann equations, written in terms of the common variable $x = \mu/T$ and the 
comoving number density $Y_{\rm DM} = n_{\rm DM}/s$ ($s$ being the entropy density), are obtained as\footnote{We adopt the notation from a recent article on two component DM \cite{Bhattacharya:2018cgx}},
\besub
\bea
\frac{dy_{H_1}}{dx} &=& \frac{-1}{x^2}\bigg{[}\langle \sigma v_{H_{1}H_{1}\rightarrow XX}\rangle \left(y_{H_1}^{2}-(y_{H_1}^{EQ})^2\right)~+~\langle \sigma v_{H_{1}H_{1}\rightarrow H_{2}H_{2}}\rangle \left ( y_{H_1}^{2}-\frac{(y_{H_1}^{EQ})^2}{(y_{H_2}^{EQ})^2} y_{H_2}^{2}\right)\Theta(m_{H_1}-m_{H_2}) \nonumber \\
&&
-~\langle\sigma v_{H_2H_2 \rightarrow H_{1}H_{1}}\rangle \left( y_{H_2}^{2}-\frac{(y_{H_2}^{EQ})^2}{(y_{H_1}^{EQ})^2}y_{H_1}^{2}\right)~\Theta(m_{H_2}-m_{H_1})\bigg{]}  =  -\frac{1}{x^2} {F_{H_1}}, \\
\frac{dy_{H_2}}{dx} &=& \frac{-1}{x^2}\bigg{[} \langle \sigma v_{H_2H_2\rightarrow XX} \rangle \left (y_{H_2}^{2}-(y_{H_2}^{EQ})^2\right )~+~\langle \sigma v_{H_2H_2\rightarrow H_{1}H_{1}}\rangle \left (y_{H_2}^{2}-\frac{(y_{H_2}^{EQ})^2}{(y_{H_1}^{EQ})^2}y_{H_1}^{2}\right )\Theta(m_{H_2}-m_{H_1}) \nonumber \\
&&
-~\langle \sigma v_{H_{1}H_{1} \rightarrow H_2H_2}\rangle \left (y_{H_1}^{2}-\frac{(y_{H_1}^{EQ})^2}{(y_{H_2}^{EQ})^2}y_{H_2}^{2}\right )\Theta(m_{H_1}-m_{H_2})\bigg{]}   = - \frac{1}{x^2} {F_{H_2}}.
\label{BE1}
\eea
\eesub
Here $y_{i}$ ($i = H_{1,2}$) is related to $Y_i$ by  $y_i=0.264M_{Pl}\sqrt{g_*}\mu Y_{i}$ and similarly for equilibrium density,
$y_i^{EQ}= 0.264M_{Pl}\sqrt{g_*}\mu Y_{i}^{EQ}$, where the equilibrium distributions are now recasted in terms of $\mu$ 
having the form
\bea
Y_{i}^{EQ}(x) = 0.145\frac{g}{g_*}x^{3/2}\bigg{(}\frac{m_{H_i}}{\mu}\bigg{)}^{3/2}e^{-x\big{(}\frac{m_{H_i}}{\mu}\big{)}}.
\eea 
Here $M_{\rm Pl} = 1.22\times10^{19} ~{\rm GeV}$ and $g_{*}=106.7$ and $X$ represents SM particles. One should note that the contribution to the Boltzmann equations coming from the DM-DM conversion (corresponding to Fig. \ref{feyn2}) will depend on the mass hierarchy of DM particles. This is described by the use of $\Theta$ function in the above equations. These coupled equations can be solved numerically to find the asymptotic abundance of the DM particles, $y_{i} \left (\frac{\mu}{m_{H_i}}x_{\infty} \right)$, which can be further used to calculate the relic:
\besub
\bea
\Omega_{i}h^2 &=& \frac{854.45\times 10^{-13}}{\sqrt{g_{*}}}\frac{m_{H_i}}{\mu}y_{H_i}\left ( \frac{\mu}{m_{H_i}}x_{\infty}\right ),
\eea
\eesub
where $x_{\infty}$ indicates a very large value of $x$ after decoupling.

In presence of neutrino Yukawa couplings, the Boltzmann equations get modified and are given by
\besub
\bea
\frac{dy_{H_1}}{dx} &=& -\frac{1}{x^2}\bigg{[} F_{H_1}+~\langle\sigma v_{H_1N_1 \rightarrow H_{2}N_{2}}\rangle 
\left (y_{H_1}y_{N_1}-\frac{y_{H_1}^{EQ}y_{N_1}^{EQ}}{y_{H_2}^{EQ}y_{N_2}^{EQ}}y_{H_2}y_{N_2} \right )~\Theta(m_{H_1}+M_1-m_{H_2}- M_2)\nonumber\\
&&
+~\langle\sigma v_{H_1N_2 \rightarrow H_{2}N_{1}}\rangle 
\left (y_{H_1}y_{N_2}-\frac{y_{H_1}^{EQ}y_{N_2}^{EQ}}{y_{H_2}^{EQ}y_{N_1}^{EQ}}y_{H_2}y_{N_1} \right )~\Theta(m_{H_1}+M_2-m_{H_2}- M_1)\nonumber\\
%
&&
~-~\langle\sigma v_{H_2N_2 \rightarrow H_{1}N_{1}}\rangle\left (y_{H_2}y_{N_2}-\frac{y_{H_2}^{EQ}y_{N_2}^{EQ}}{y_{H_1}^{EQ}y_{N_1}^{EQ}}y_{H_1}y_{N_1}\right )~\Theta(m_{H_2}+M_2-m_{H_1}-M_1)\nonumber \\
&&
~-~\langle\sigma v_{H_2N_1 \rightarrow H_{1}N_{2}}\rangle\left (y_{H_2}y_{N_1}-\frac{y_{H_2}^{EQ}y_{N_1}^{EQ}}{y_{H_1}^{EQ}y_{N_2}^{EQ}}y_{H_1}y_{N_2}\right )~\Theta(m_{H_2}+M_1-m_{H_1}-M_2)\bigg{]}, \\
\frac{dy_{H_2}}{dx} &=& -\frac{1}{x^2}\bigg{[} F_{H_2}~+~\langle\sigma v_{H_2N_2 \rightarrow H_{1}N_{1}}\rangle \left(y_{H_2}y_{N_2}-\frac{y_{H_2}^{EQ}y_{N_2}^{EQ}}{y_{H_1}^{EQ}y_{N_1}^{EQ}}y_{H_1}y_{N_1}\right)~\Theta(m_{H_2}+M_2-m_{H_1}- M_1)\nonumber\\
&&
+~\langle\sigma v_{H_2N_1 \rightarrow H_{1}N_{2}}\rangle \left(y_{H_2}y_{N_1}-\frac{y_{H_2}^{EQ}y_{N_1}^{EQ}}{y_{H_1}^{EQ}y_{N_2}^{EQ}}y_{H_1}y_{N_2} \right)~\Theta(m_{H_2}+M_1-m_{H_1}- M_2)\nonumber\\
%
&&
~-\langle\sigma v_{H_1N_1 \rightarrow H_{2}N_{2}}\rangle \left(y_{H_1}y_{N_1}-\frac{y_{H_1}^{EQ}y_{N_1}^{EQ}}{y_{H_2}^{EQ}y_{N_2}^{EQ}}y_{H_2}y_{N_2} \right)~\Theta(m_{H_1}+M_1-m_{H_2}- M_2)\nonumber\\
&&
~-\langle\sigma v_{H_1N_2 \rightarrow H_{2}N_{1}}\rangle \left(y_{H_1}y_{N_2}-\frac{y_{H_1}^{EQ}y_{N_2}^{EQ}}{y_{H_2}^{EQ}y_{N_1}^{EQ}}y_{H_2}y_{N_1} \right)~\Theta(m_{H_1}+M_2-m_{H_2}- M_1)\bigg{]}.
\eea
\label{BE2}
\eesub
The total relic density of DM follows from the combined contribution of both the components and is given by 
$\Omega_{\rm{total}} = \Omega_{H_1} h^2 + \Omega_{H_2} h^2$. 

\subsection{\textbf{Direct Detection}}
\begin{figure}[H]
\centering
\includegraphics[scale=0.50]{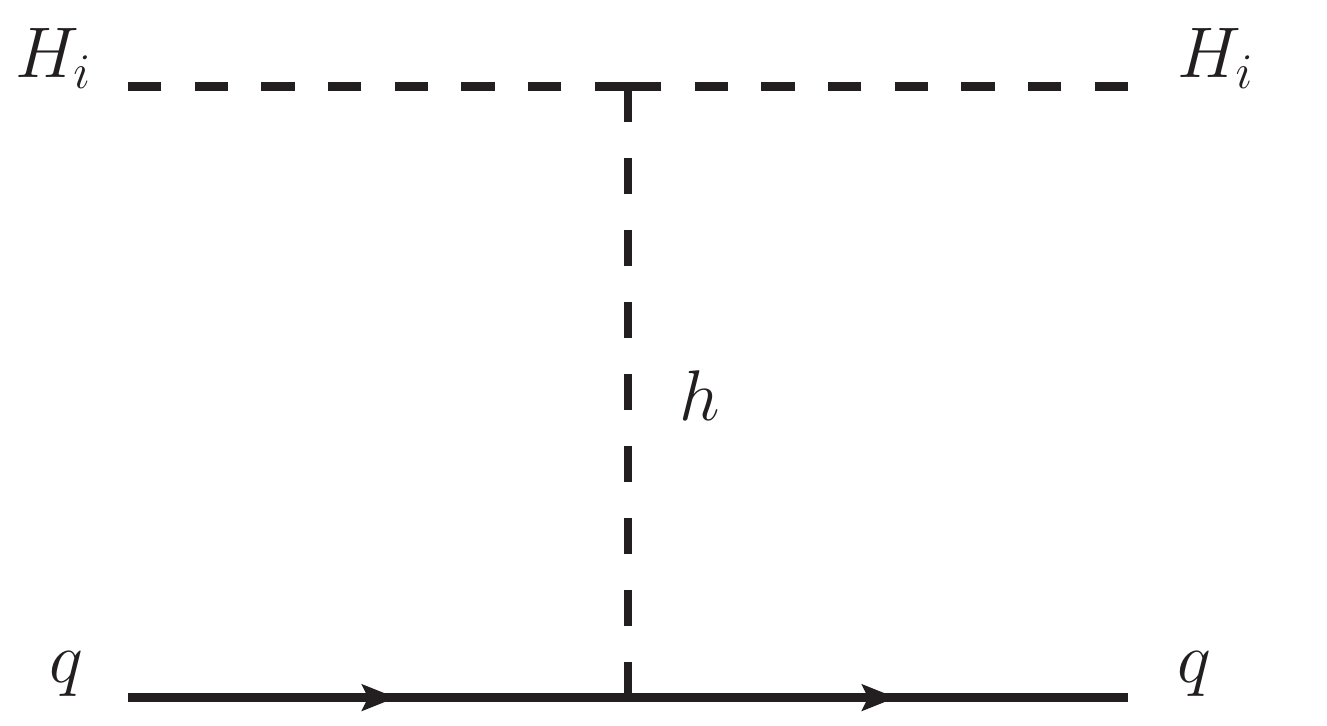}
\caption{Spin independent elastic scattering of DM-nucleon.}
\label{feyndd1}
\end{figure}
As mentioned earlier, DM parameter space can be constrained significantly by the null results at different direct detection experiments such as LUX \cite{Akerib:2016vxi}, PandaX-II \cite{Tan:2016zwf, Cui:2017nnn} and Xenon1T \cite{Aprile:2017iyp, Aprile:2018dbl}. There are two ways scalar DM can scatter off nuclei at tree level in our model. One is elastic scattering 
mediated by SM Higgs boson while the other one is the inelastic one mediated by electroweak gauge bosons. The latter 
can be kinematically forbidden by considering large mass splitting between IHD components, typically larger than the 
average kinetic energy of DM particle. The spin independent elastic scattering cross section mediated by SM Higgs 
(shown in Fig. \ref{feyndd1}) is given as \cite{Barbieri:2006dq}
\begin{equation}
 \sigma^{\text{SI}}_i = \frac{\lambda^2_L f^2_n}{4\pi}\frac{\mu_{i,n}^2 m^2_n}{m^4_h m^2_{i}},
\label{sigma_dd}
\end{equation}
where $\mu_{i,n} = m_n m_{i}/(m_n+m_{i})$ is the ${\rm DM}$-nucleon reduced mass and $\lambda_{L_i}$ is the quartic 
coupling involved in ${\rm DM}$-Higgs interaction. The index $i$ stands for DM candidate in our scenario: $H_1, ~H_2$. 
A recent estimate of the Higgs-nucleon coupling $f$ gives $f = 0.32$ \cite{Giedt:2009mr} although the full 
range of allowed values is $f=0.26-0.63$ \cite{Mambrini:2011ik}. In this two-component DM framework, 
the spin-independent cross section relevant for each of the candidate can be expressed as 
\begin{equation}
\sigma^{\text{SI}}_{\rm{i, ~eff}} = \frac{\Omega_{H_i}}{\Omega_{\rm{total}}}\sigma^{\text{SI}}_{H_i}.
\label{sigma-eff}
\end{equation} 
Latest results from Xenon-1T experiment provides a strong constraint on single component IDM as 
it restricts $\l_L$ coupling significantly. However due to the presence of two DM components here 
in our set-up, such tight constraints can be evaded by suitable adjustment of relative DM abundance. 
We will discuss the status of our model at direct detection frontier in subsequent sections.

\subsection{\textbf{Indirect Detection}}
As mentioned earlier, WIMP DM candidates have good prospects at indirect detection experiments looking for excess of gamma rays. DM particles can annihilate and produce SM particles, out of which photons (and also neutrinos), being electromagnetically neutral, have better chances of reaching the detector from source without getting deflected. Following the notations of \cite{Ahnen:2016qkx}, the observed differential gamma ray flux produced due to the DM annihilation can be computed as 
\begin{equation}
\frac{d\Phi}{dE} (\triangle \Omega) = \frac{1}{8\pi} \langle \sigma v \rangle  \frac{J (\triangle \Omega)}{M^2_{\text{DM}}} \frac{dN}{dE},
\end{equation}
where $\triangle \Omega$ is the solid angle corresponding to the observed region of the sky, $\langle \sigma v \rangle$ is the thermally averaged DM annihilation cross section, $dN/dE$ is the average gamma ray spectrum per annihilation process and the astrophysical $J$ factor is given by
\begin{equation}
J(\triangle \Omega) = \int_{\triangle \Omega} d\Omega' \int_{\rm LOS} dl \rho^2(l, \Omega').
\end{equation}
In the above expression, $\rho$ is the DM density and LOS corresponds to line of sight. Therefore, measuring the gamma ray 
flux and using the standard astrophysical inputs, one can constrain the DM annihilation first into different charged final states 
like $\mu^+ \mu^-, \tau^+ \tau^-, W^+ W^-, b\bar{b}$ which in turn produces the gamma rays. As discussed in case of 
direct detection, here also our set-up carries some flexibility as far as indirect detection constraints are concerned. We 
incorporate the global analysis of the Fermi-LAT and MAGIC observations of dSphs \cite{Ahnen:2016qkx} for this purpose. 
The bounds quoted in \cite{Ahnen:2016qkx} consider $100\%$ annihilation of DM into particular final states as well as 
assume a single DM component which fills the entire $26\%$ of the universe. Since our construction involves deviation from these 
consideration, we can make the bounds weaker by playing with the branching fraction to a particular final states and 
simultaneously changing the relative fractional abundance. This is because the DM annihilation rates are directly proportional to number density squared of DM in local neighbourhood.

\section{Neutrino Mass}
\label{sec3}

\begin{figure}[H]
\centering
\includegraphics[scale=0.50]{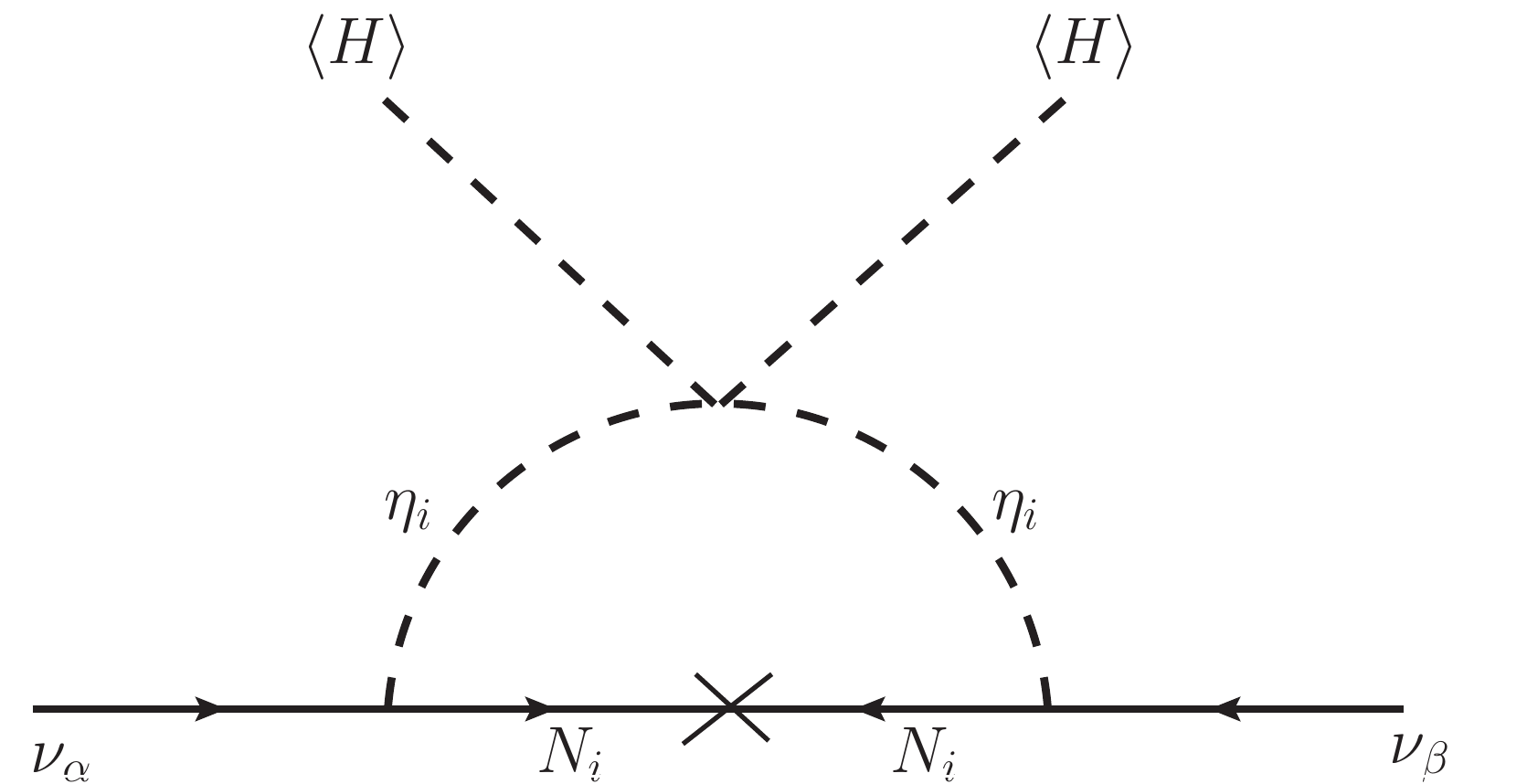}
\caption{Radiative generation of light neutrino mass.}
\label{feynnumass}
\end{figure}
Similar to the minimal scotogenic model, here also light neutrino masses are generated at one loop level, as shown in Fig. \ref{feynnumass}. However, since each $\mathbb{Z}_2$ sector contains only one singlet fermion and one IHD, combination 
of one loop contribution from both $\mathbb{Z}_2 \times \mathbb{Z'}_2 $ sectors leads to the light neutrino mass matrix 
elements as  
\bea
(m^{\nu})_{\alpha\beta}&=&\sum_{i=1,2}\frac{Y_{\alpha i}Y_{\beta i}M_{i}}{
{32}\pi^2}\bigg{[}\frac{m_{H_{i}}^2}{m_{H_{i}}^2-M_{i}^2} {\rm{ln}}\frac{m_{H_{i}}^{2}}{
M_{i}^{2}}-\frac{m_{A_i}^2}{m_{A_i}^2-M_{i}^2} {\rm{ln}}\frac{m_{A_i}^{2}}{M_{i}^{2}} 
\bigg{]},\\
&=& \sum_{i=1,2} Y_{\alpha i}~\Lambda_{ii}~Y^{T}_{i \beta},
\label{nu1}
\eea
where $m_{H_i}$ and $m_{A_i}$ are provided in Eq. (\ref{eq5}). Expressions of $\Lambda_{ii}$ is then 
found to be 
\bea
\Lambda_{ii}&=&\bigg{[}\frac{M_i}{
{32}\pi^2}\bigg{(}\frac{m_{H_i}^2}{m_{H_i}^2-M_{\color{red}i}^2} {\rm{ln}}\frac{m_{H_i}^{2}}{
M_i^{2}}-\frac{m_{A_i}^2}{m_{A_i}^2-M_i^2} {\rm{ln}}\frac{m_{A_i}^{2}}{M_i^{2}} 
\bigg{)}\bigg{]}_{ii}.
\label{nu2}
\eea

The light neutrino mass matrix $m_{\nu}$ can be diagonalised through 
\begin{equation}
m_{\nu} = U_{\rm PMNS} m^{\rm diag}_{\nu} U^T_{\rm PMNS},
\end{equation} 
where $m^{\rm diag}_{\nu} = {\text{diag}}(m_{{\nu}_1}, m_{{\nu}_2}, m_{{\nu}_3})$.
Considering the charged lepton mass matrix as a diagonal one, the light neutrino diagonalising matrix $U_{\text{PMNS}}$ 
coincides with the lepton mixing matrix.  The $U_{\text{PMNS}}$ matrix can be parametrised as 
\begin{equation}
U_{\text{PMNS}}=\left(\begin{array}{ccc}
c_{12}c_{13}& s_{12}c_{13}& s_{13}e^{-i\delta}\\
-s_{12}c_{23}-c_{12}s_{23}s_{13}e^{i\delta}& c_{12}c_{23}-s_{12}s_{23}s_{13}e^{i\delta} & s_{23}c_{13} \\
s_{12}s_{23}-c_{12}c_{23}s_{13}e^{i\delta} & -c_{12}s_{23}-s_{12}c_{23}s_{13}e^{i\delta}& c_{23}c_{13}
\end{array}\right) U_{\text{P}},
\label{matrixPMNS}
\end{equation}
where $c_{ij} = \cos{\theta_{ij}}, \; s_{ij} = \sin{\theta_{ij}}$ and $\delta$ is the leptonic Dirac CP phase. 
The diagonal phase matrix $U_{\text{P}}=\text{diag}(1, e^{i\alpha}, e^{i\beta})$  contains the Majorana CP phases 
$\alpha, \beta$ which remain undetermined at neutrino oscillation experiments. We summarise the $3\sigma$ global 
fit values in Table \ref{tabglobalfit} from the recent global fit \cite{Esteban:2018azc}, which we use in our 
subsequent analysis.

\begin{table}[htb]
\centering
\begin{tabular}{|c|c|c|}
\hline
Parameters & Normal Hierarchy (NH) & Inverted Hierarchy (IH) \\
\hline
$ \frac{\Delta m_{21}^2}{10^{-5} \text{eV}^2}$ & $6.79-8.01$ & $6.79-8.01 $ \\
$ \frac{|\Delta m_{31}^2|}{10^{-3} \text{eV}^2}$ & $2.427-2.625$ & $2.412-2.611 $ \\
$ \sin^2\theta_{12} $ &  $0.275-0.350 $ & $0.275-0.350 $ \\
$ \sin^2\theta_{23} $ & $0.418-0.627$ &  $0.423-0.629 $ \\
$\sin^2\theta_{13} $ & $0.02045-0.02439$ & $0.02068-0.02463 $ \\
$ \delta (^\circ) $ & $125-392$ & $196-360$ \\
\hline
\end{tabular}
\caption{Global fit $3\sigma$ values of neutrino oscillation parameters \cite{Esteban:2018azc}.}
\label{tabglobalfit}
\end{table}

Note that the two $\mathbb{Z}_2$ sectors in our model can generate at most two light neutrino masses\footnote{With the involvement of a third RH neutrino to start with and making it very heavy, it effectively leads to the two RH neutrino scenario 
we consider here. As a consequence of this limit, it is shown \cite{Antusch:2011nz} that one of the light neutrino remains massless.}, which needs to 
be consistent with light neutrino data mentioned above. This leaves us with two possibilities (a) $m_1=0, m_2 < m_3$, 
(b) $m_1 < m_2, m_3=0$ corresponding to normal and inverted hierarchies respectively. 
Since the inputs from neutrino data are only in terms of the mass squared differences and mixing angles (phases are considered to be zero here), it would be  
useful for our purpose to express the Yukawa couplings in terms of light neutrino parameters. This is possible through the 
Casas-Ibarra (CI) parametrisation \cite{Casas:2001sr} extended to radiative seesaw model \cite{Toma:2013zsa} which 
allows us to write the Yukawa couplings as
\bea
Y&=&U_{\text{PMNS}}\sqrt{m^{diag}_{\nu}}R^{\dagger}\sqrt{\Lambda^{-1}}, 
\label{nu5}
\eea
where $\Lambda$ is the 2$\times$2 diagonal matrix with eigenvalues defined in Eq.(\ref{nu2}) and $R$ is a complex 
orthogonal matrix \cite{Ibarra:2003xp, Ibarra:2003up, Antusch:2011nz} having the form,
\bea
R &=&
  \begin{pmatrix}
    0 & \cos{z} & \sin{z}\\
    0 & -\sin{z} & \cos{z}\\
  \end{pmatrix}.
\eea
Using Eq.(\ref{nu5}), the elements of $3 \times 2$ Yukawa matrix can be obtained with specific choices of the 
complex angle $z$. 
The same calculation can be repeated for inverted hierarchy as well. In the subsequent sections, we 
discuss how the constraints from neutrino sector can play a non-trivial role in the dark matter parameter 
space from relic abundance criteria.

\section{Lepton Flavour Violation}
\label{sec31}
Since the charged lepton flavour violating (LFV) decays are very much suppressed in the SM, any observation of such effects 
will be a clear signature of beyond the SM physics. In our model, due to the coupling of each $\mathbb{Z}_2$ sector particles 
($N_i$ and $\eta_i$) to the SM leptons, one may expect some contribution to such LFV effects at one loop level. The same 
fields that take part in the one-loop generation of light neutrino mass as shown in Fig. \ref{feynnumass} also mediate 
LFV processes like $\mu \rightarrow e \gamma$. The neutral scalar in the internal lines of Fig. \ref{feynnumass} will 
be replaced by their charged counterparts (which emit a photon) whereas the external fermion legs can be replaced by $\mu, 
e$ respectively, generating the one-loop contribution to $\mu \rightarrow e \gamma$. 

As the couplings and masses involved in this process are the same as the ones that generate light neutrino masses and play 
a role in DM relic abundance, we can no longer choose them arbitrarily. It should be noted that each $\mathbb{Z}_2$ sector contributes separately to this process and hence we have to add the respective contributions at amplitude level. Adopting the notations from \cite{Baek:2014awa}, we can write 
\bea
{\rm Br}(L_{\alpha}\rightarrow L_{\beta}\gamma)&=&\frac{3\alpha_{\rm em}}{64\pi G_{F}^2}\bigg{|}\sum_{i=1,2} \frac{Y_{\beta i}Y_{\alpha i}^{*}}{m^2_{\eta^{+}_{i}}}F\bigg{(}\frac{M_i^2}{m_{\eta^{+}_{i}}^2}\bigg{)}\bigg{|}^2,
\label{LFV1}
\eea
Here $\alpha_{\rm em}=e^2/4\pi$ is the electromagnetic fine structure constant, $G_F$ is the Fermi constant, $F(x)$ = $(1-6x+3x^2+2x^3-6x^2 \log{x})/(6(1-x)^4)$ is the loop factor and $x=\frac{M_i^2}{m_{\eta_{i}^+}^2}$. The MEG experiment provides the most stringent upper limit on the branching ratio: ${\rm Br}(\mu \rightarrow e \gamma) < 4.2 \times 10^{-13}$ \cite{TheMEG:2016wtm}.

\section{Results and Discussion}
\label{sec4}
As mentioned earlier, we write the model in \texttt{LanHEP} \cite{Semenov:2014rea} and then extract the model files to 
use in \texttt{micrOMEGAs 4.1} \cite{Belanger:2014vza} for two component DM framework. To have a better understanding 
of the role of neutrino Yukawa coupling on the phenomenology of DM, we first discuss the results (i) without any involvement 
of neutrino Yukawa interactions and then (ii) incorporate the Yukawa interactions. Finally constraints from light neutrino data 
and LFV decays are studied in order to constrain the parameter space. 

\subsection{Results in the absence of neutrino Yukawa coupling}

As mentioned earlier in the introduction, one of the key motivations for studying the two component scalar doublet dark matter  
over other existing possibilities is related to the existence of an intermediate region of a single scalar doublet dark matter mass 
$80 \; \text{GeV} \leq M_{\rm DM} \leq 500 \; \text{GeV}$  where correct relic abundance can not be satisfied irrespective of 
the choices of other parameters. While in the single component DM framework, this intermediate region can be revived by 
additional production mechanisms \cite{Molinaro:2014lfa, Borah:2017dfn}, there still exists severe bounds from indirect detection experiments which effectively rule out single scalar doublet DM mass below around 400 GeV \cite{Borah:2017dfn}. Reviving 
this intermediate region by incorporating an additional DM component in the form of a fermion and vector boson have been 
studied recently by the authors of \cite{Chakraborti:2018lso} and \cite{Chakraborti:2018aae} respectively. In this work, we consider the 
other remaining possibility of introducing a second scalar doublet DM component which carries non-trivial implications for 
light neutrino masses (which most of the multi-component DM models do not address) also. Since intermediate mass region is 
special in a sense that one component scalar doublet DM is not enough to produce correct relic density, we stick to that mass 
range (roughly) for our numerical analysis.

\begin{figure}[H]
\centering
\subfigure[]{
\includegraphics[scale=0.50]{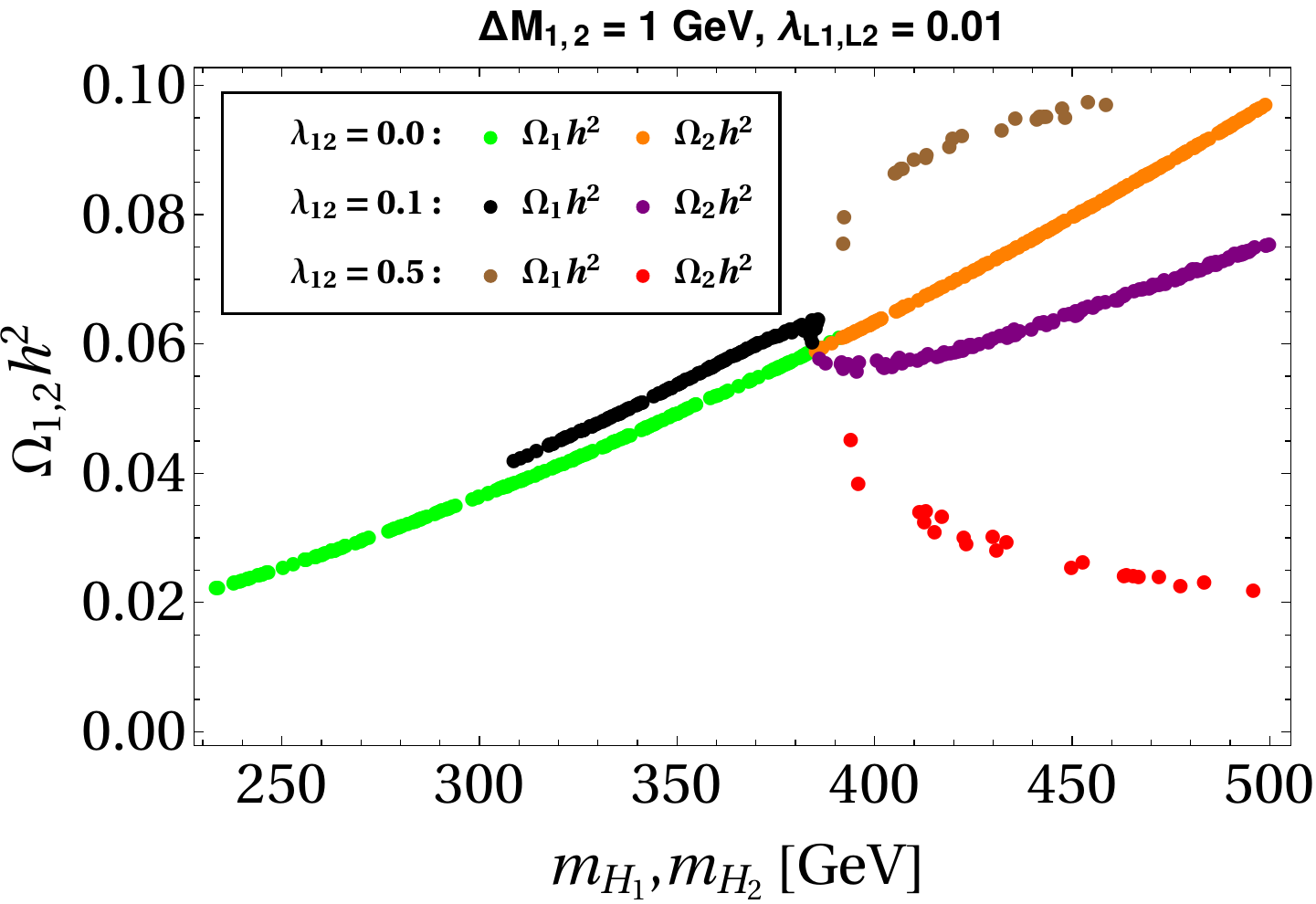}}
\subfigure[]{
\includegraphics[scale=0.50]{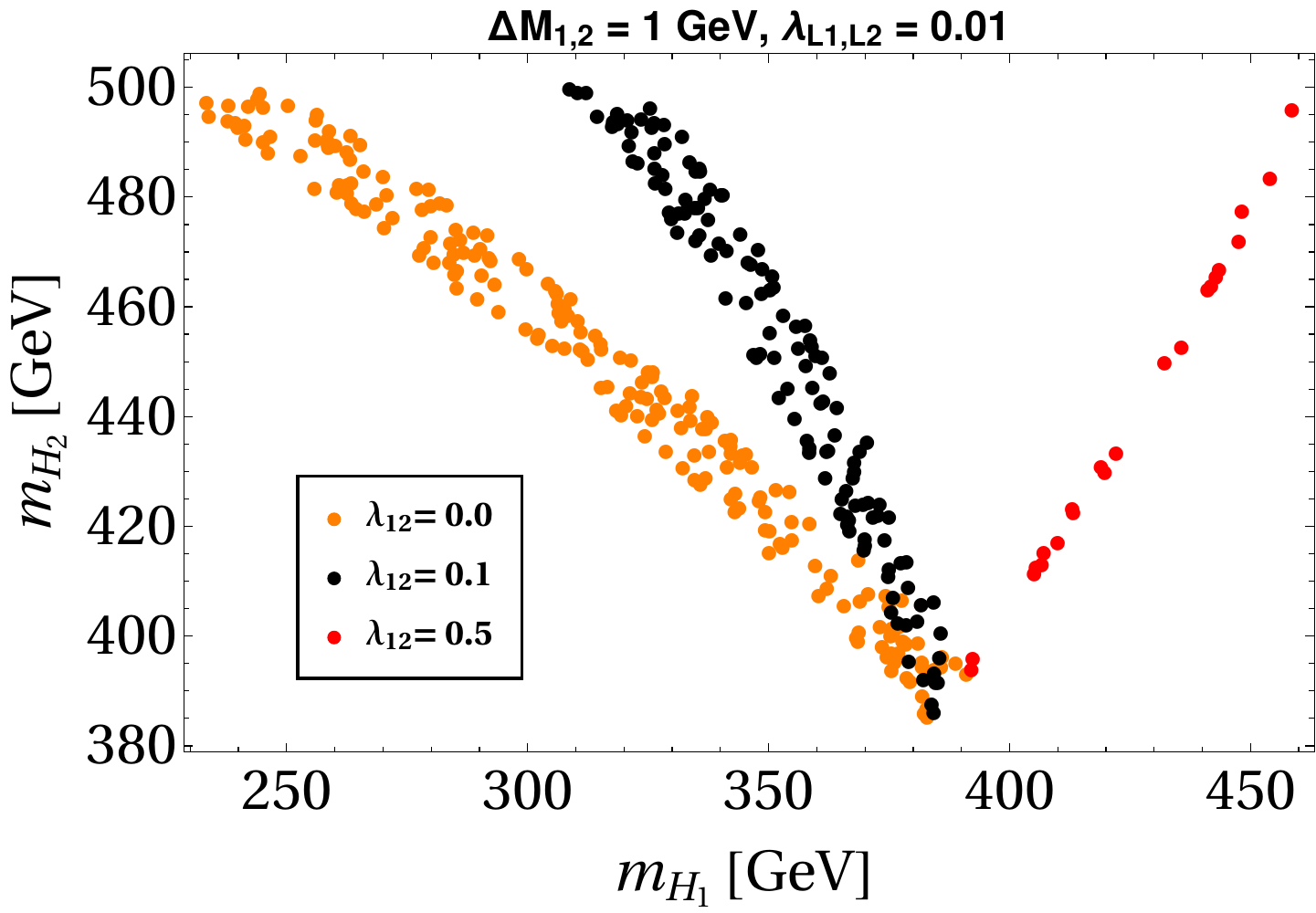}}
\caption{Points which satisfy the correct total DM relic abundance and are allowed by the direct detection constraints for 
different values of $\l_{12}$  while maintaining $m_{H_2}>m_{H_1}$.}
\label{fig4}
\end{figure}

The left panel of Fig. \ref{fig4} shows the variation of individual contribution to the relics $\Omega_{1(2)} h^2$ against 
the mass of dark matter $m_{H_{1(2)}}$ such that $\Omega_{\rm total}h^2$ satisfies the Planck 2018 limit. The same 
variation for different values of the conversion coupling $\lambda_{12}$ is shown with three different choices of 
$\lambda_{12}$ =  0 (green for $H_1$ and orange for $H_2$ contribution), 0.1 (black for $H_1$ and purple for 
$H_2$ contribution), 0.5 (brown for $H_1$ and red for $H_2$ contribution). Furthermore, we consider $m_{H_2} 
> m_{H_1}$ within our range of interest of DM mass. The other relevant parameters are kept fixed at their respective 
benchmark values: mass splitting within $\eta_1 (\eta_2), ~i.e.~\Delta M_{i=1,2} = m_{A_i} - m_{H_i} =1$ GeV and 
$\l_{L_{1,2}} = 0.01$. This mass splitting plays role in individual coannihilation diagrams, similar to single component IDM. 
Such choice of $\Delta M_i$ is motivated from the fact that the maximum contribution from a single DM component 
toward relic density can obtained for such small mass difference.  $\l_{L_{1,2}}$ values are chosen for representative 
purpose. 

It is intriguing to note also from Fig. \ref{fig4} (both from left and right panel) that a change of pattern of the 
patches of different colours happened around $m_{H_1} = m_{H_2} \simeq 380$ GeV and that small region (where 
all these points meet) is almost independent of the change of $\l_{12}$ value.  At the meeting point, where 
both the DM masses are around 380 GeV, almost $50\%$ individual contribution follows from both the DM candidates. 
Since $H_{1,2}$ are degenerate at this point, no number changing process can take place at this point irrespective 
of the fact whether $\l_{12}$ is zero or non-zero. Hence it explains why all the patches meet over this region. 
Note that while obtaining these plots, we keep the IHD masses within the stipulated intermediate range: 80-500 GeV. 
Finally what we observe from the plots is that the mass range 230-500 GeV becomes allowed range while incorporating 
a second inert doublet. 

Fig. \ref{fig4} is clearly an indicative of the importance of the parameter $\l_{12}$, which plays a non-trivial role in 
conversion of one DM candidate into the other. In order to understand this, let us begin with $\l_{12}~=~0$ case, 
as shown with green and orange patches. When $\Omega_{1} h^2$ is small, $\Omega_{2} h^2$ should be large such that 
$\Omega_{\rm total}h^2$ can satisfy the Planck 2018 limit. This indicates that a point (say $m_{H_1}$=320 GeV with 
$\Omega_1 h^2$ = 0.0402 ) in the lower side of green patch ($i.e.$ with low $m_{H_1}$) is actually correlated to a single  
point (say $m_{H_2}$= 445 GeV with $\Omega_2 h^2$ = 0.0789) near the higher side orange patch (with relatively large 
$m_{H_2}$). Effect of the conversion coupling $\l_{12}$ would be clear with the plot in right panel of Fig. \ref {fig4}, 
where the total relic satisfied points are placed in $m_{H_1}, m_{H_2}$ plane. Given a fixed $m_{H_1}$, required masses 
of $m_{H_2}$ are shown (hence the code of $H_2$ are given as used in left panel) with $m_{H_2} > m_{H_1}$. 
Note that for $m_{H_1} = 320$ GeV,  the relic constraint would be satisfied by $m_{H_2}$ =  491 GeV with 
$\l_{12} = 0.1$. Hence it is clear that due to conversion $H_2 H_2 \rightarrow H_1 H_1$, somewhat higher 
value (as compared to that of $\l_{12}$ = 0) of $m_{H_2}$ becomes allowed. One should keep in mind that 
here the combination of couplings $\l^{\prime}_3$, $\l^{\prime}_4$, $\l^{\prime}_5$ appears in the conversion 
processes via $\l_{12}$ and hence even for the smaller values of individual couplings, large conversion effects can 
be seen. 

\begin{figure}[H]
\centering
\subfigure[]{
\includegraphics[scale=0.50]{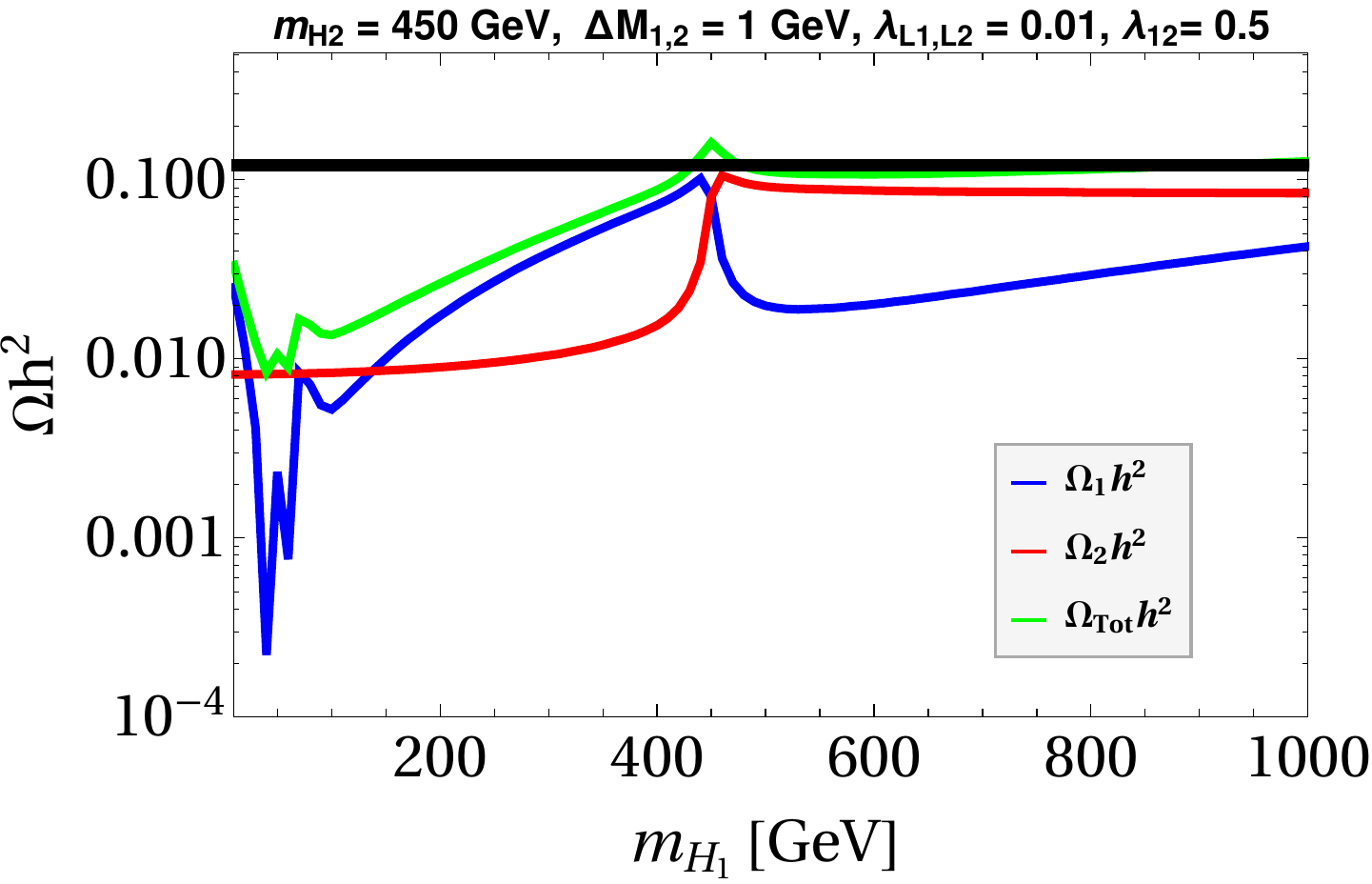}}
\subfigure[]{
\includegraphics[scale=0.50]{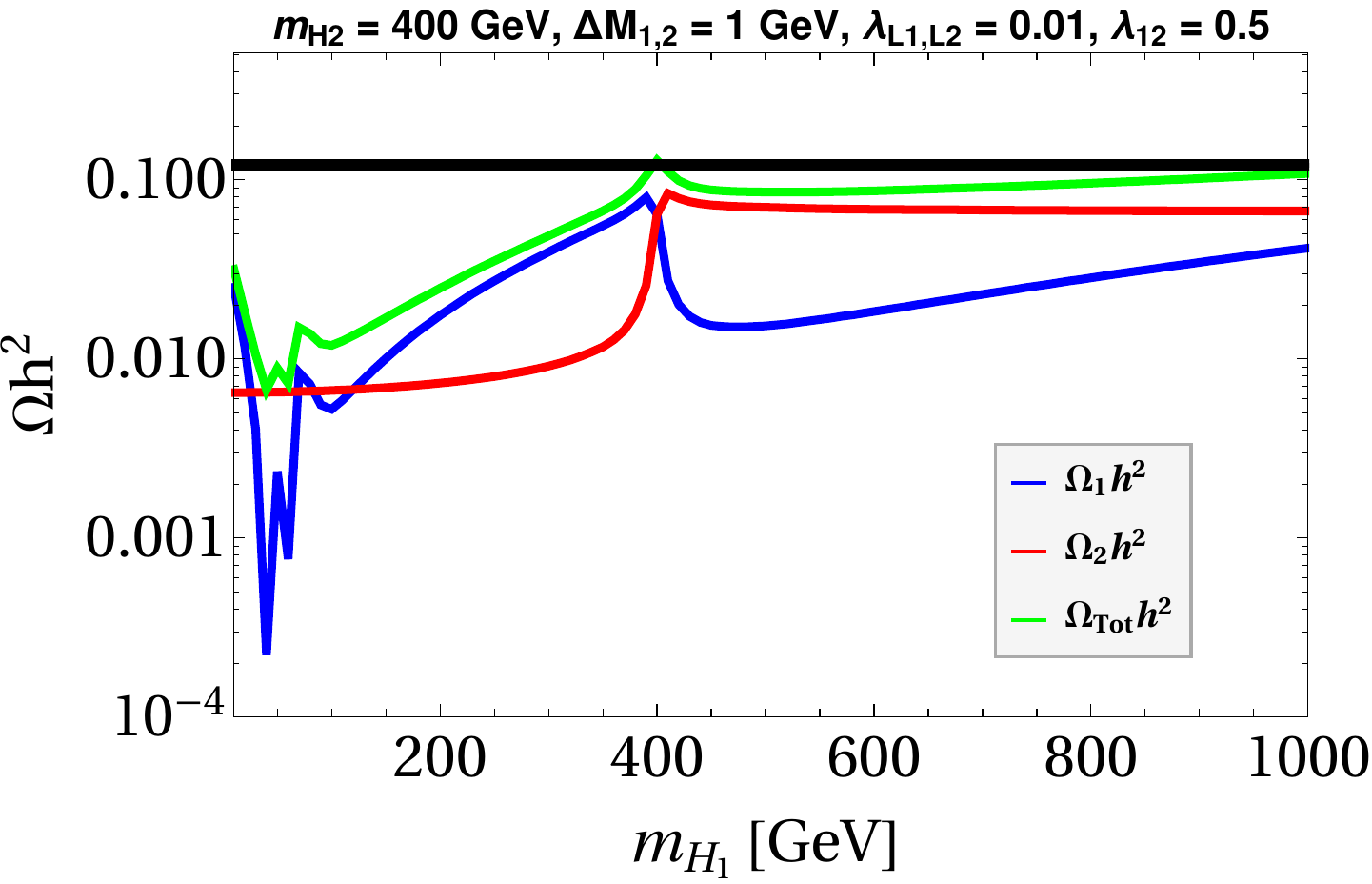}}
\caption{Relic abundance of DM candidates as a function of $H_1$ mass for fixed benchmark values of other parameters.}
\label{fig3}
\end{figure}
Also note that once we are away from this specific point, a typical pattern (parabolic behaviour) of the plots sets 
in. In order to have a better understanding of it, we plot the variation of individual and total relic abundance with 
DM masses for fixed values of conversion parameter $\l_{12} =0.5$ and $m_{H_2}$ = 450 GeV, while keeping 
other parameters to their benchmark values. The results are shown in Fig. \ref{fig3}. 
For $m_{H_1}$ below 450 GeV, conversion like $H_2 H_2 \rightarrow H_1 H_1$ is effective. As the mass of $H_1$ 
crosses the threshold of $H_2$ mass (here 450 GeV), the conversion ($H_1 H_1 \rightarrow H_2 H_2$) becomes 
efficient thereby reducing the abundance of $H_1$ while increasing the latter's. The total relic curve (green line) is 
almost symmetric around this cross-over point which explains the typical nature of the left panel plot of Fig. \ref{fig4} 
in the high mass regime with large conversion coupling.  

\begin{figure}[H]
\centering
\subfigure[]{
\includegraphics[scale=0.50]{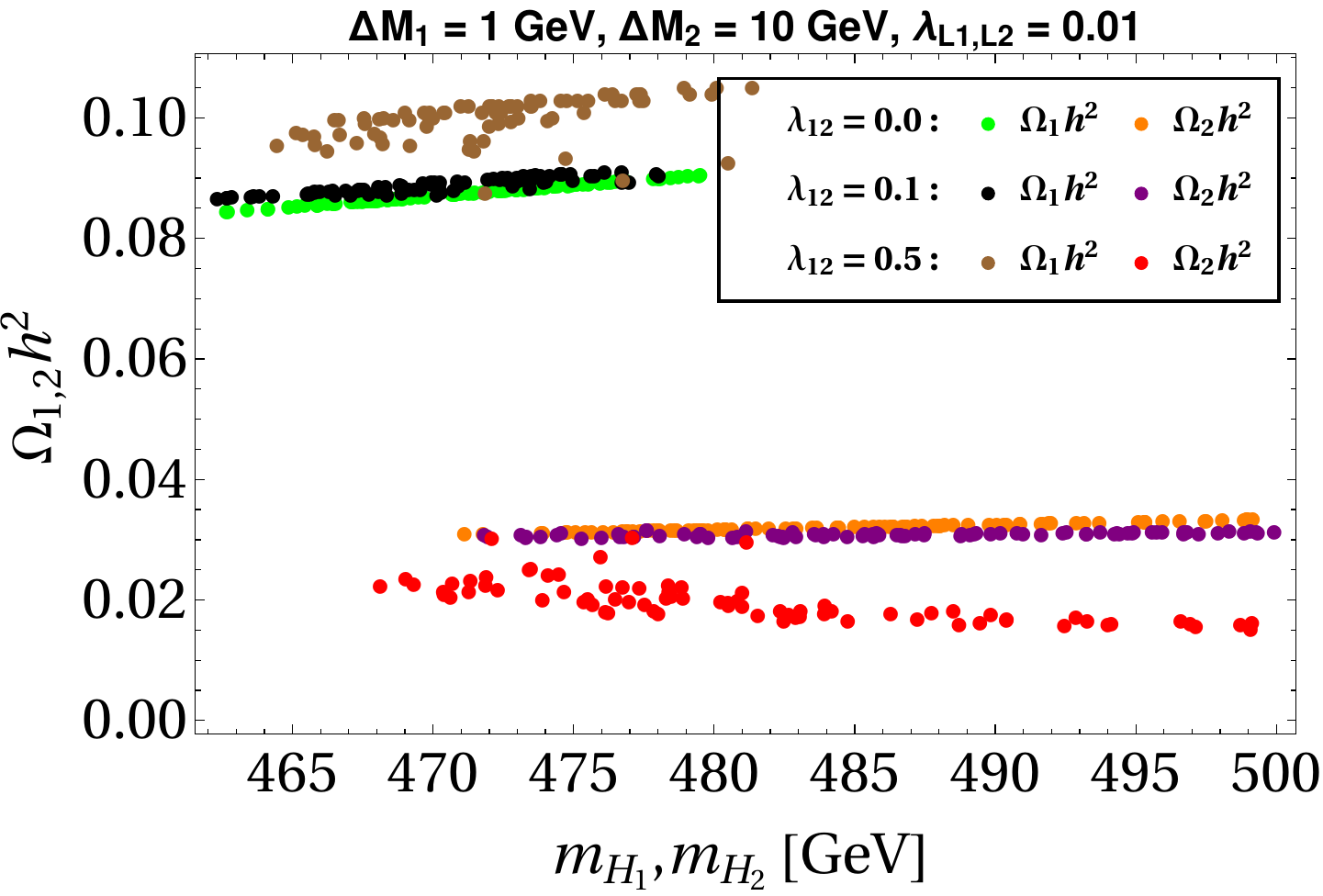}}
\subfigure[]{
\includegraphics[scale=0.50]{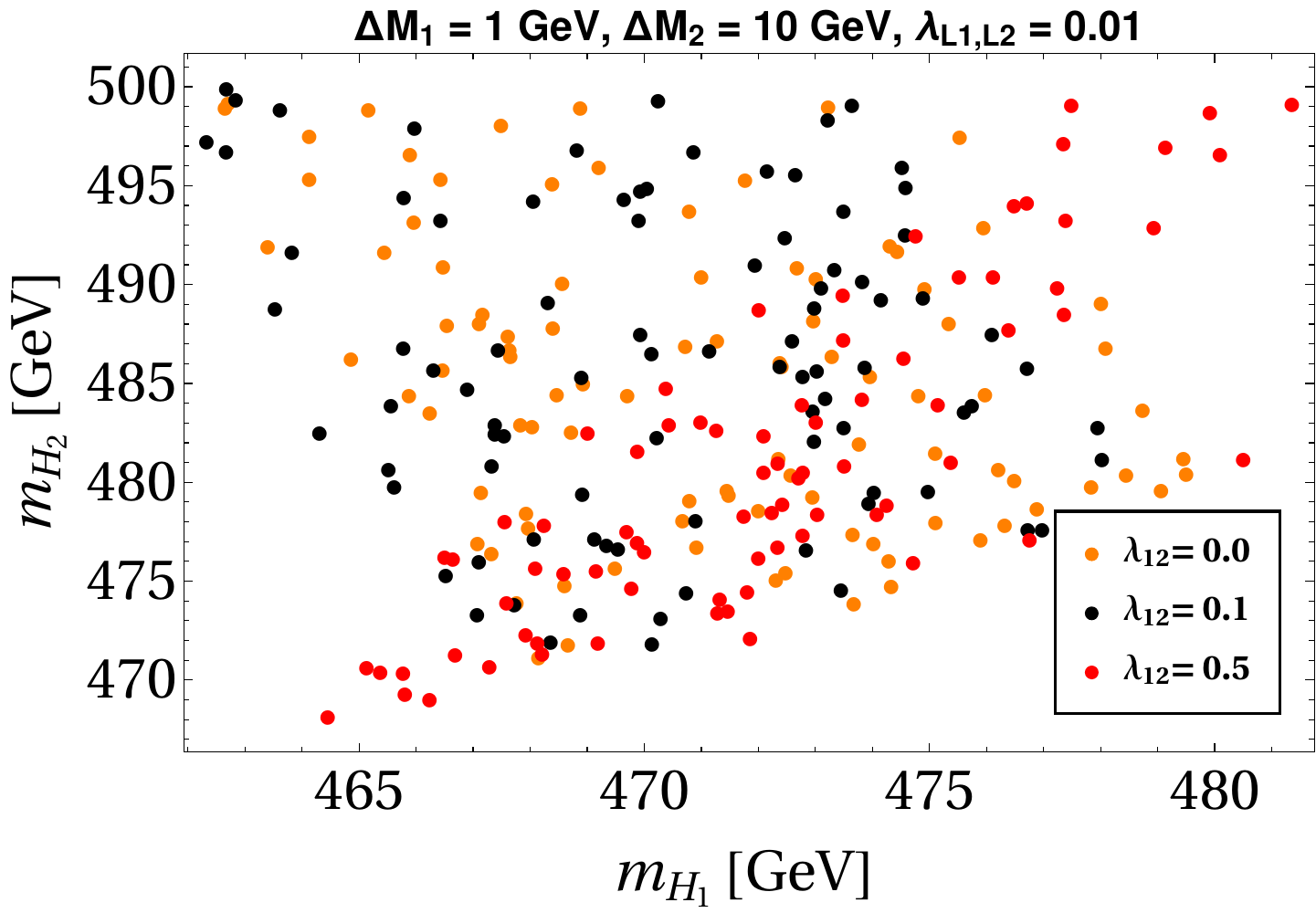}}
\caption{Points which satisfies the correct total DM relic abundance and are allowed by the direct detection constraints for different values of $\l_{12}$ for $\Delta M_{2}=10$ GeV while maintaining $m_{H_2}>m_{H_1}$.}
\label{fig6}
\end{figure}
We also check the effects of considering different mass splittings $\Delta M_1 \neq \Delta M_2$. As we increase 
$\Delta M_2$ to 10 GeV while keeping $\Delta M_1$ at 1 GeV, the relic abundance of $H_2$ decreases, a feature 
observed in single component IDM also in the high mass regime. The results are shown in Fig. \ref{fig6} (considering 
$m_{H_2}>m_{H_1}$) which shows that the masses of $H_2$ are shifted to higher mass regions in order to satisfy 
the total DM relic abundance. This is expected as we know based on our knowledge of single component inert doublet DM 
analysis that making the mass splitting related to one IHD more, the corresponding relic density would be less. This 
clarifies why there is a separation between the green and orange patches (with $\l_{12} = 0$). Similar situation prevails 
when nonzero $\l_{12}$ value is switched on as well, which can be seen from black (brown) and purple (red) patches. 
Also notice that the existence of the symmetric point similar to the case with equal $\Delta M_i$ (about which the relic 
contour lines gain a typical shape with non-zero $\l_{12}$) is lost here. This is related to the fact that now two components 
of DM have different type of co-annihilations and hence contribute to the total relic differently. This is also reflected 
from a mixed up distribution pattern of correct relic satisfied points put in $m_{H_1}, m_{H_2}$ plane in the right panel of 
Fig. \ref{fig6}. One more point to notice is that there is a shift toward the higher mass range of DM as compared to the 
case displayed in Fig. \ref{fig4} with $m_{H_2} > m_{H_1}$.

\begin{figure}[H]
\centering
\subfigure[]{
\includegraphics[scale=0.50]{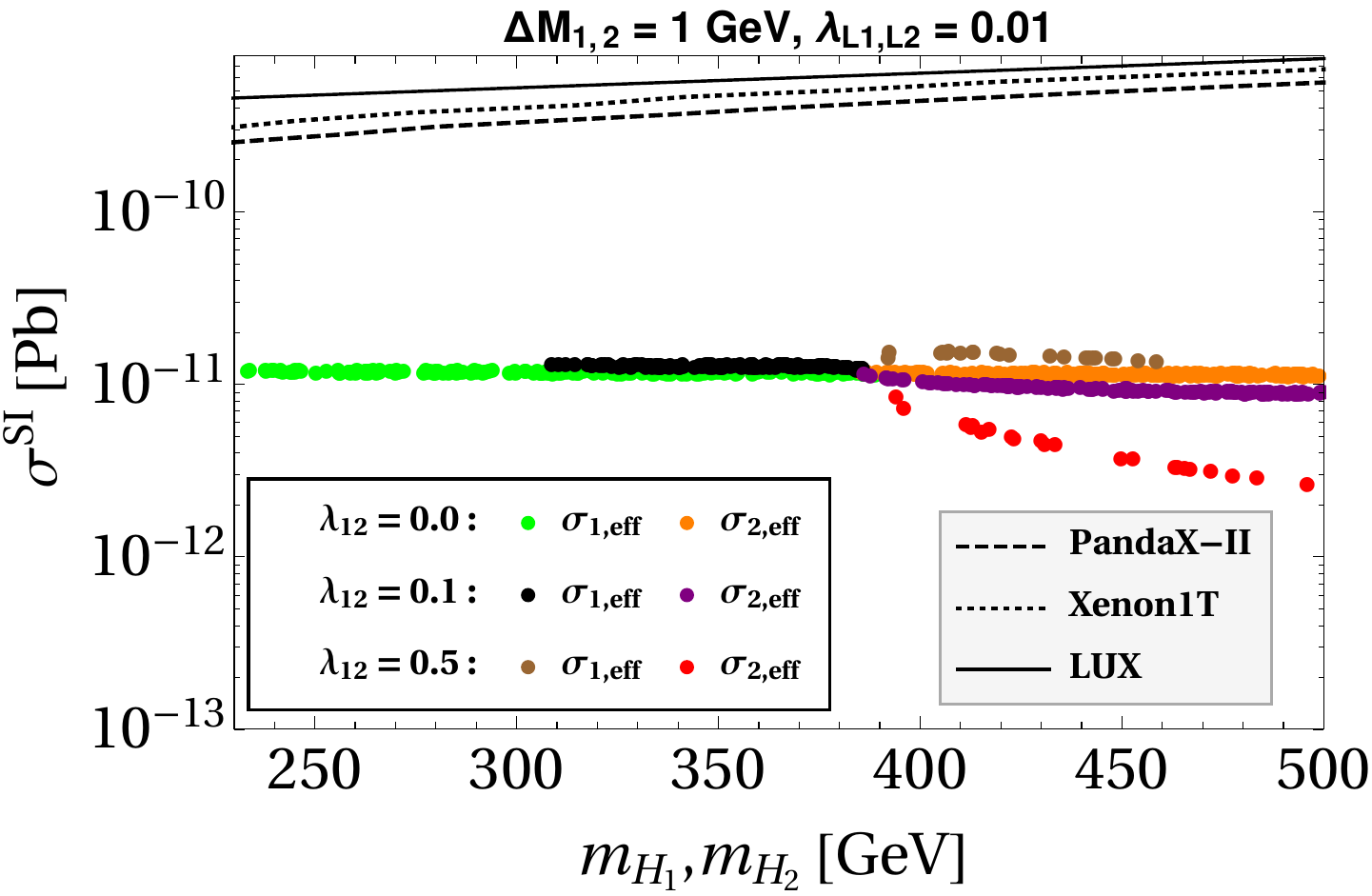}}
\subfigure[]{
\includegraphics[scale=0.50]{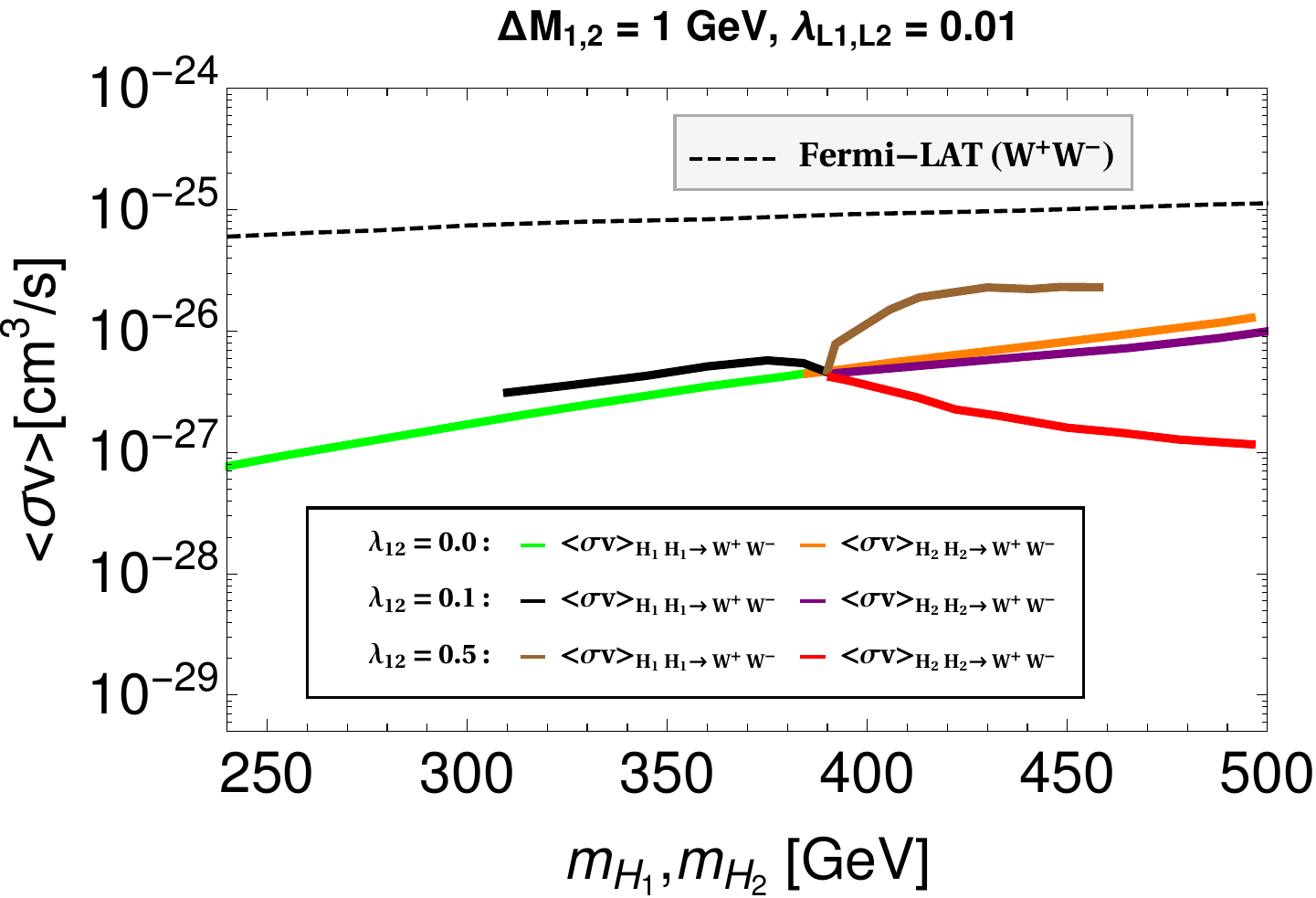}}
\caption{(a) Left panel: Spin independent DM-nucleon scattering cross section from our model ($m_{H_2}>m_{H_1}$) 
for different choices of $\l_{12}$. Limits from direct detection experiments are shown in black  dashed (PandaX-II), dotted (Xenon1T) 
and solid lines (LUX); (b) Right panel: Indirect detection cross section for $W^+ W^-$ final states with different benchmark 
parameters satisfying the correct relic and direct detection bounds ($m_{H_2}>m_{H_1}$). Upper bounds from Fermi-LAT experiment are denoted 
by black dashed line.}
\label{D-ID}
\end{figure}

As mentioned earlier, scalar doublet DM has good prospects for detection at direct search experiments. This is due to 
the presence of gauge as well as Higgs portal interactions with nucleons. While the gauge portal interactions at tree 
level can be forbidden kinematically, the Higgs portal interactions can still give rise to sizeable spin independent elastic 
scattering of DM off nucleons. For representative purpose of our model, we fix the Higgs portal coupling $\l_{L_{1,2}}$ at 
0.01 for both the DM candidates and provide their direct detection cross section for different conversion parameters 
$\l_{12}$ in the left panel of Fig. \ref{D-ID}. Note that we include those points only which satisfy the total DM relic 
abundance. We have taken into account the relative abundance of individual DM candidates while calculating the 
direct detection cross section. The colour code used here is consistent to the discussion above related to Fig. {\ref{fig4}}.
As can be seen from this plot, the scattering cross sections in our model remains well below the latest direct search bounds. However, with increase in the value of Higgs portal coupling, these points will be closer to the experimental upper bound 
and hence providing the model a good prospect of being detected at ongoing and near future experiments like LZ~\cite{Akerib:2015cja}, XENONnT~\cite{Aprile:2015uzo}, DARWIN~\cite{Aalbers:2016jon} and PandaX-30T~\cite{Liu:2017drf}.\\
One should note that it will be difficult to distinguish the direct detection signals coming from both the DM candidates in reality. As pointed out in \cite{Herrero-Garcia:2017vrl} it is possible to distinguish the direct detection signals for a two-component scenario from that of the single-component dark matter provided the two dark matter masses differ significantly (by an order of magnitude or more). In our present set-up, both the dark matter masses are with in the intermediate mass range ($80-500$) GeV and hence the possiblity is no more useful here.

We also check the prospects for indirect detection of DM in our model by specifically focussing on $W^+ W^-$ final states from DM annihilations. This is due to the chosen mass range of DM where annihilation to this final states is the most dominant one. We incorporate the relative abundance of the two DM candidates while calculating their annihilation rates. We include the factor coming from the branching fraction to this final state from DM annihilations. Our findings are displayed in the right panel of 
Fig. {\ref{D-ID}}. It can be seen that for some part of the parameter space, the indirect detection cross section lies 
close to the experimental upper bound. This plot also points out the increase in allowed parameter space for two component 
inert doublet DM compared to one component inert DM where DM mass below 400 GeV was ruled out by Fermi-LAT constraints \cite{Borah:2017dfn}.

\subsection{Results in the presence of neutrino Yukawa couplings}
After discussing the new results for two component inert DM with salient features arising due to gauge and Higgs portal 
interactions along with four point interactions, we now move on to discuss the role of neutrino Yukawa interactions with  
two-component DM candidates. This has non-trivial connections to the light neutrino sector as the same Yukawa couplings 
also play a role in generating light neutrino masses. The new annihilation channels that will come into play now are the 
ones shown in Fig. \ref{feyn1a}. We now have an additional free parameter which is important for dark matter phenomenology
defined as $\Delta M_{NiHi}~=~M_i-m_{H_i}$. For analysis purpose, we fix it to $\Delta M_{NiHi}~=~10~{\rm GeV}$ value. 
Such small mass splitting between neutral singlet fermion and DM candidates enhances the co-annihilation cross section 
effectively having significant effect on the relic abundance as we will see soon.

In obtaining the relic density and direct/indirect detection cross sections in our model, we vary the IHD masses within 
the so called intermediate range of DM mass. Since we have chosen $\Delta M_{NiHi}~=~10~{\rm GeV}$, this can 
automatically fix the two RH neutrino masses for chosen values of $m_{H_1, H_2}$. Hence with the benchmark values 
(as specified in last subsection) of $\Delta M_{1,2}$, we would get the elements of $\Lambda$ matrix. Then using 
Eq. (\ref{nu5}), one can determine the neutrino Yukawa couplings for some choices of $z$ involved in $R$ matrix.  
We have made the simplest choice by setting $z$ = 0 in order to determine $R$ matrix. The light neutrino masses are evaluated from the best 
fit values of solar and atmosphere mass-square splittings as one of the light neutrino remains massless in the scenario.

\begin{figure}[H]
\centering
\subfigure[]{
\includegraphics[scale=0.50]{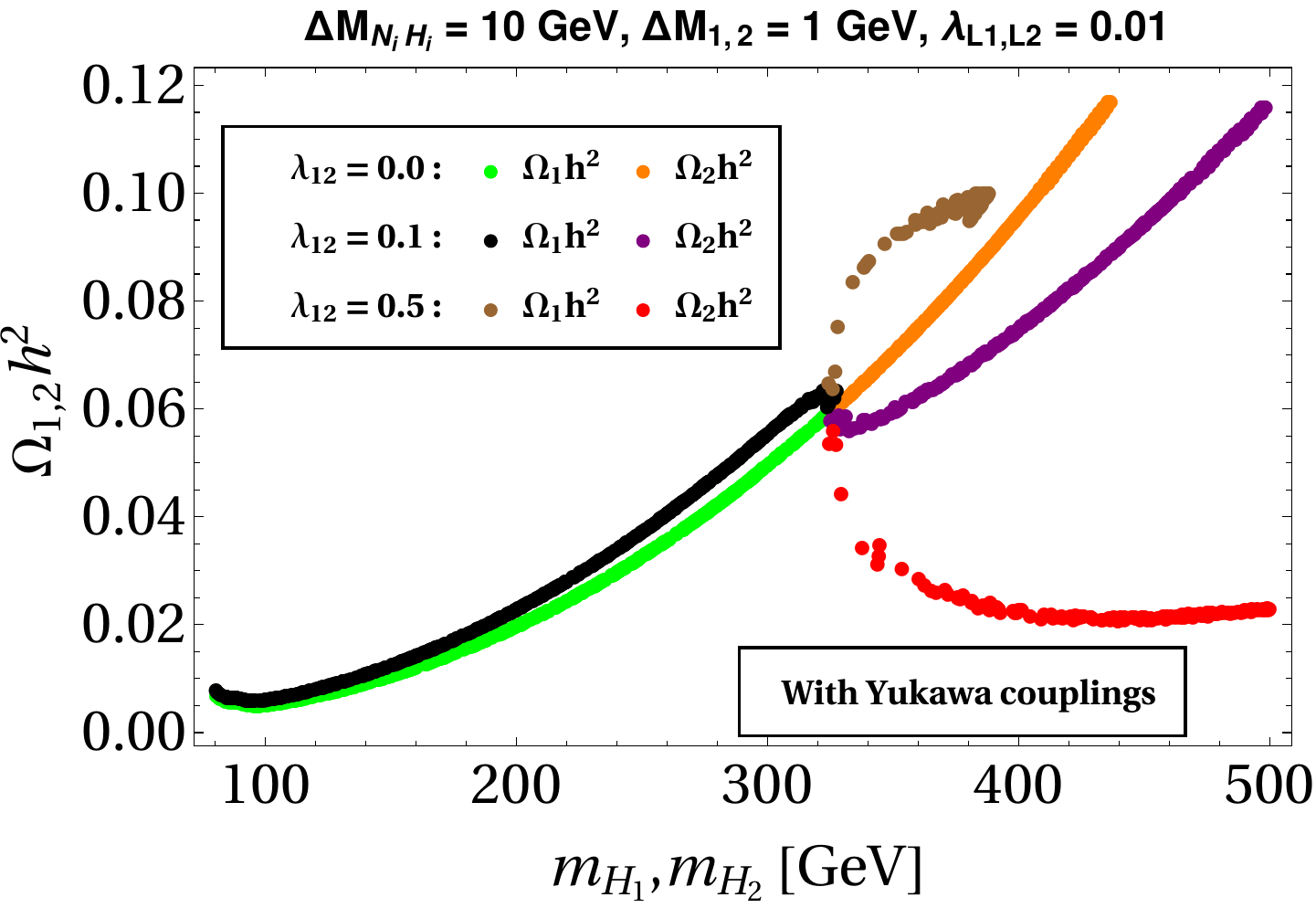}}
\subfigure[]{
\includegraphics[scale=0.50]{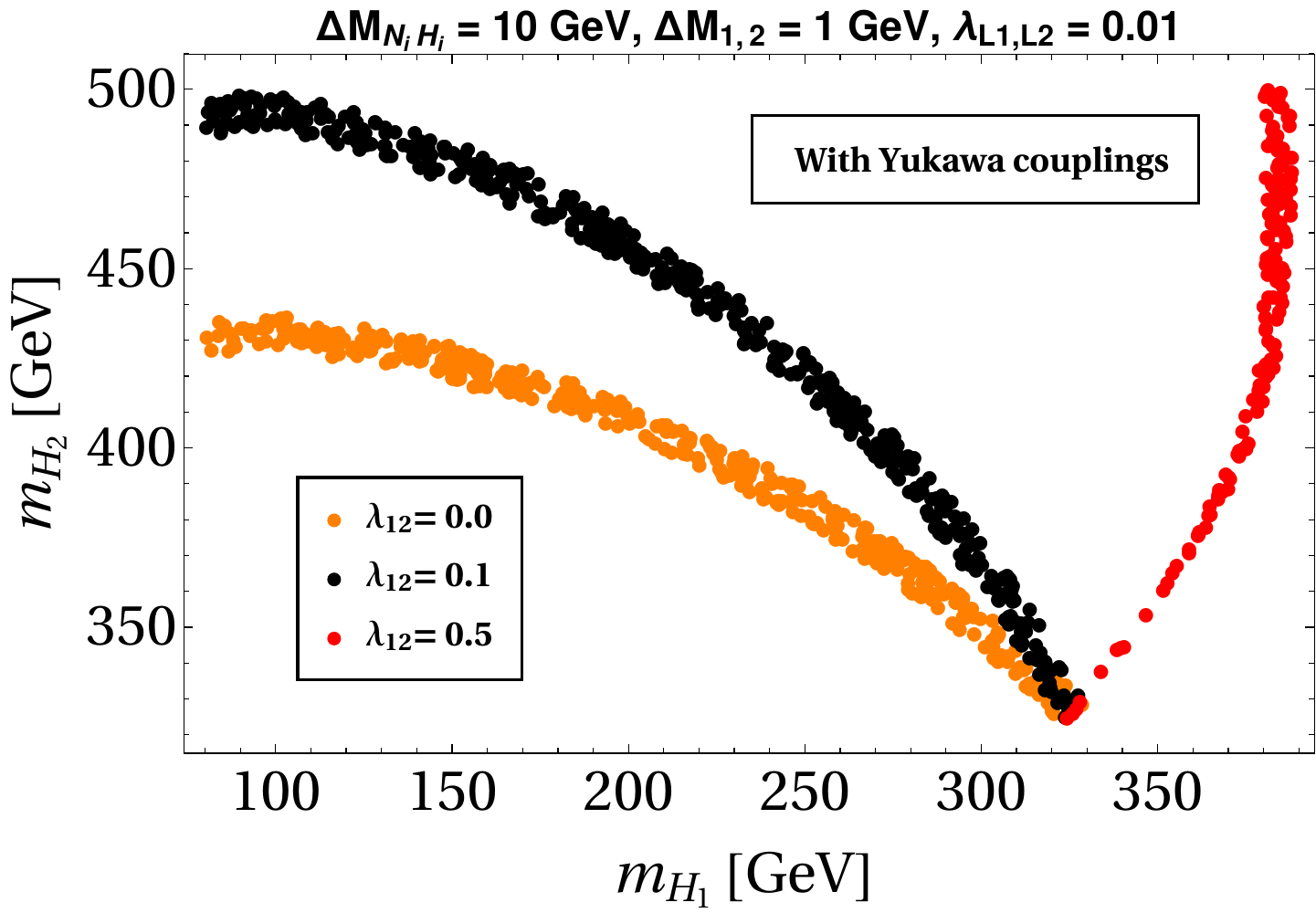}}
\caption{Points which satisfies the correct total DM relic abundance and are allowed by the direct detection constraints for different values of $\l_{12}$ while maintaining $m_{H_2}>m_{H_1}$. An analogue of the plots in Fig. \ref{fig4} in the presence of Yukawa interactions. }
\label{fig8}
\end{figure}

\begin{figure}[H]
\centering
\includegraphics[scale=0.70]{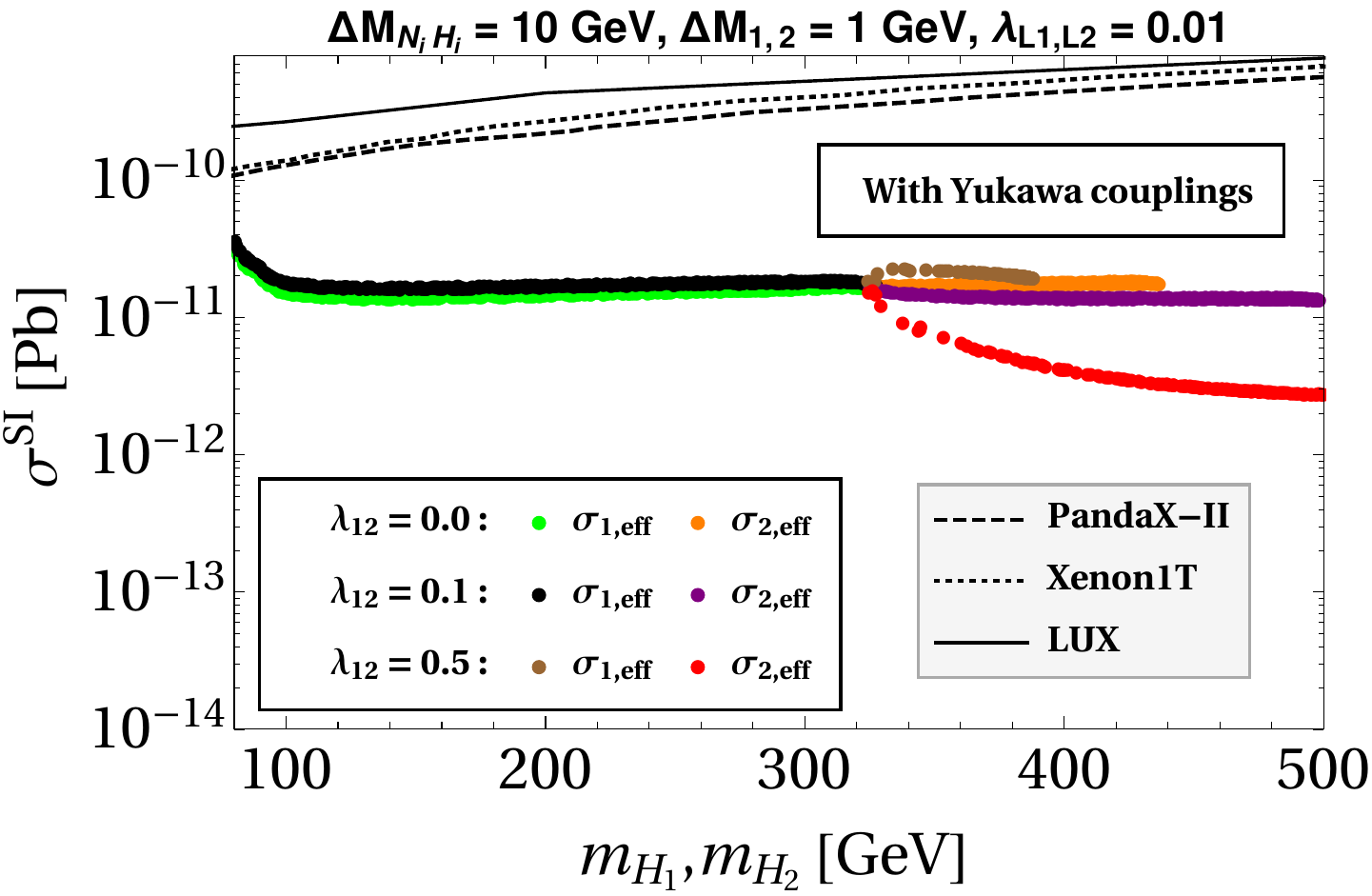}
\caption{Spin independent DM-nucleon scattering cross section as compared to upper limits from direct detection experiments. All points satisfy the total DM relic criteria and $m_{H_2}>m_{H_1}$. An analogue of the plot shown in Fig. \ref{D-ID} in the presence of Yukawa interactions.}
\label{DD2}
\end{figure}

We first reproduce the parameter space which gives rise to the total DM relic abundance in agreement with Planck 
2018 limit and also allowed by direct detection bounds for different benchmark values of conversion coupling $\l_{12}$. The individual contributions of $H_1$ and $H_2$ to the total relic are displayed in Fig. \ref{fig8}(a). 
The same results, in terms of the relic contours in $m_{H_1}$-$m_{H_2}$ plane (while maintaining $m_{H_2}>m_{H_1}$), are 
shown in Fig. \ref{fig8}(b) which if compared with the one in the absence of Yukawa interactions (see Fig. \ref{fig4}(b)) reveals 
interesting shift in parameter space. For example with $\Delta M_{1,2}$ = 1 GeV, the symmetric point (near $m_{H_1} 
\sim 380$ GeV) about which patterns of relic contours corresponding to different $\l_{12}$ values evolve in Fig. \ref{fig4} 
now shifted to 330 GeV. As this point corresponds to $m_{H_1} = m_{H_2}$ having no contribution followed from 
co-annihilations involving DM components, new co-annihilations channels involving neutrino Yukawa couplings 
(see Fig. \ref{feyn1a}) would anyway be present. This causes the observed shift.\\
It turns out that the required Yukawa couplings are of order $10^{-4}-10^{-5}$ for both normal and inverted hierarchies. This estimate is obtained from $m_{H_1}-m_{H_2}$ contours (allowed by both the relic and direct detection constratints)\footnote{We have checked that with the inverted hierarchy, the results in terms of the values of parameters involved, do not change significantly.} along with the expression in Eq.(\ref{nu5}) with the use of Eq.(\ref{nu2}).  
The other interesting observation is: incorporating Yukawa interactions allow smaller values of DM masses that satisfy 
correct total DM abundance and hence the entire intermediate region for single inert doublet DM becomes allowed with 
two-component inert DM model involving neutrinos. As stated above, the presence of the Yukawa couplings would introduce additional co-annihilation channels (see Fig. \ref{feyn1a}). While Fig. \ref{feyn1a}(a) and (b) contribute to both the DMs in a similar way, the DM-DM conversions (via Fig. \ref{feyn1a}(c) and (d))   can effectively alter the two DM components differently (depending on their mass hierarchy) as seen in the Boltzmann equation via Eq.(\ref{BE2}a) and Eq.(\ref{BE2}b). So with $m_{H_2}>m_{H_1}$, in general the processes $H_2N_1\rightarrow H_1N_2$ (with non-zero $\l_{12}$) and  $H_2N_2\rightarrow H_1N_1$ would enhance the number density of $H_1$ component \footnote{With the choice of $\Delta M_{N_iH_i}=10$ GeV only $H_2N_2\rightarrow H_1N_1$ would contribute.} and hence its contribution towards relic ($\Omega_1h^2$) becomes more while compared with a similar $m_{H_1}$ having no Yukawa interactions. This can be seen by comparing Fig.\ref{fig4}(a) and Fig. \ref{fig8}(a). For $m_{H_1}=250$ GeV, $\Omega_1h^2$ was 0.026 as seen from Fig. \ref{fig4}(a). Now for same $m_{H_1}$, but with Yukawa interaction, $\Omega_1h^2$ becomes 0.033. Accordingly $m_{H_2}$ is shifted toward a lower mass value to satisfy the total relic to be consistent with observation. In this particular example, $m_{H_2}$ comes down to 380 GeV (Fig. \ref{fig8}(a)) from its earlier value of 492 GeV (Fig. \ref{fig4}(a)).
Although the Yukawa interactions are not supposed to change 
the direct and indirect detection rates at tree level, due to the change in the prefactor $\frac{\Omega_{H_i}}{\Omega_{total}}$ of $\sigma^{SI}_{H_i}$, some changes in $\sigma^{SI}_{i,eff}$ are expected. This is shown in  Fig. \ref{DD2}.

\begin{figure}[H]
\centering
\includegraphics[scale=0.50]{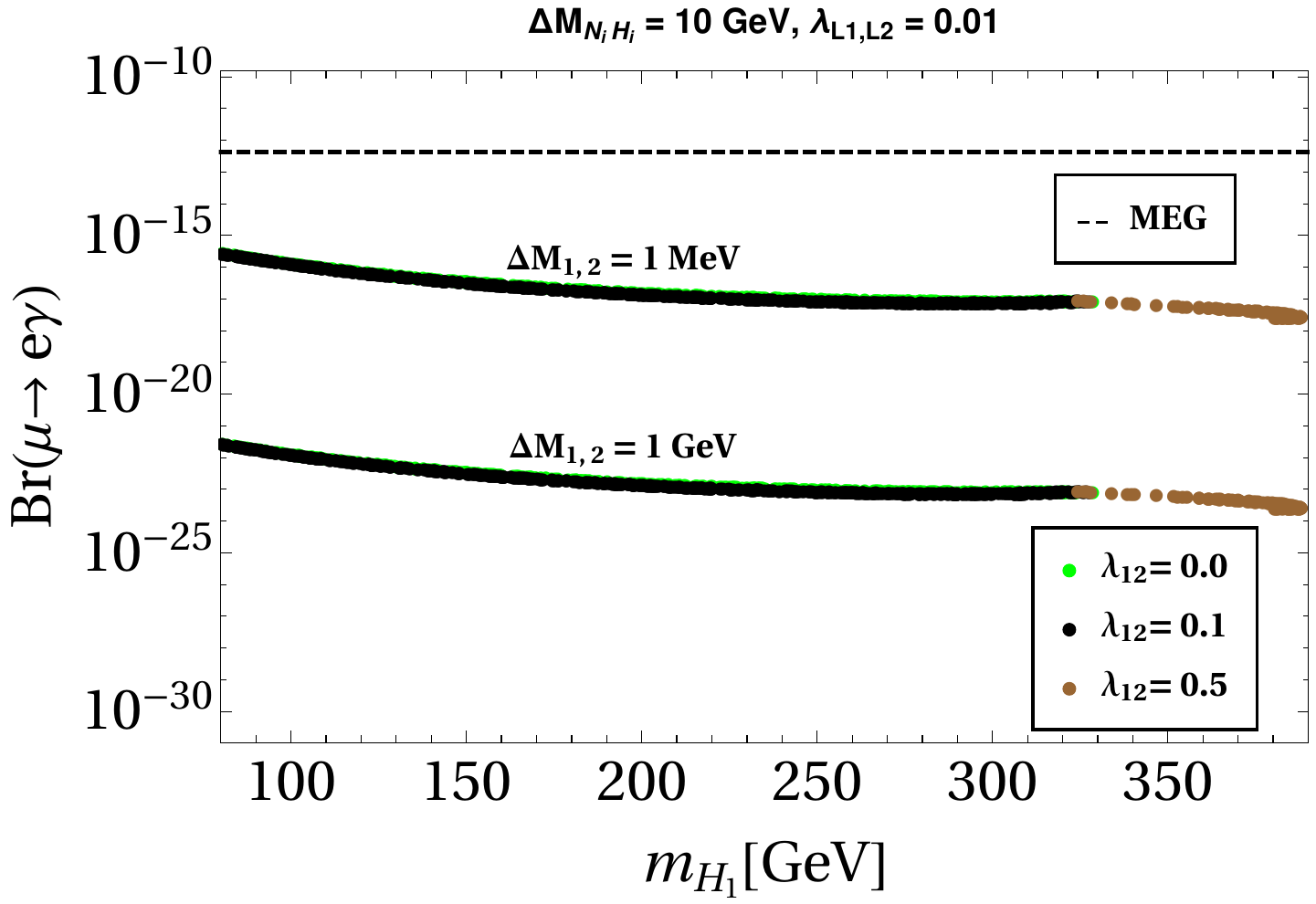}
\caption{Contributions to ${\rm Br}(\mu\rightarrow e \gamma)$ as a function of DM mass $m_{H_1}$}
\label{fig9}
\end{figure}

We finally show the predictions for LFV decay, $\mu \rightarrow e \gamma$, in Fig. \ref{fig9} by choosing similar benchmark 
values of mass splittings as in the dark matter analysis (used in Fig. \ref{fig8}) against variation with $m_{H_1}$. 
In obtaining this plot, we have used that  pair of $m_{H_1}, m_{H_2}$ values which satisfy relic density constraint with their total contribution. The lower set of plots is for $\Delta M_{i} = 1$ GeV while upper set is with $\Delta M_{i} = 1$ MeV. Note that 
we have carried out the DM phenomenology with  $\Delta M_{i} = 1$ GeV only. If we work with further small 
values of mass-splitting, the DM phenomenology would not change, particularly when we concentrate our interest within the intermediate range of DM mass: 80 GeV - 500 GeV. However while making both of these splittings smaller, it will increase branching 
ratio 
related to the prediction of LFV decays as seen from the Fig. {\ref{fig9}}. This is particularly due to the increase in Yukawa couplings for smaller values of mass splitting, in accordance with the CI parametrisation. Hence with not-so-small values of the mass 
splittings, we can conclude that the range of the parameter space allowed by the DM relic density, direct and indirect search results and neutrino data is mostly unaffected by the LFV constraints. 
Finally it is perhaps pertinent to discuss the collider propects of the present scenario. Due to the involvement of the associated charged components of the two IHDs, the typical opposite sign di-lepton plus missing energy signature can be quite interesting at LHC \cite{Bhattacharya:2019fgs}. On top of that, in our present model, both the DM masses are within the intermediate range ($<500$ GeV). Hence it would be more promising in our present set-up in terms of detection at LHC compared to the IHD DM present in the usual scotogenic scenario\cite{Ma:2006km}. We provide the production cross-section for $pp\rightarrow H^+H^-$ in the following Table \ref{tab3} which was generated in CalcHEP \cite{Belyaev:2012qa} at $\sqrt{s}=14$ TeV using cteq6l1 \cite{Placakyte:2011az} parton distribution corresponding to a benchmark scenario with $m_{H_1(H_2)}=300 (350)~\rm{GeV}$. The respective relic density and the direct detection cross-section are also indicated in the same table.
%

\begin{table}[H]
\centering
\begin{tabular}{|c|c|c|c|c|c|c|c|c|c|c|}
\hline 
\multirow{2}{*}{BP} & \multicolumn{4}{c|}{Masses of (in GeV)}
 & \multirow{2}{*}{$\Omega_{1}h^2$} & \multirow{2}{*}{$\Omega_{2}h^2$} & \multirow{2}{*}{ $\sigma_{1,eff}$[pb]} & \multirow{2}{*}{$\sigma_{2,eff}$[pb]} & \multirow{2}{*}{$\sigma_{pp\rightarrow H_{1}^{+}H_{1}^{-}}$[fb]} & \multirow{2}{*}{$\sigma_{pp\rightarrow H_{2}^{+}H_{2}^{-}}$[fb]} \\
 \cline{2-5}
 &$H_1$ & $H_1^{+}$ & $H_2$ & $H_2^{+}$ & & & & & &\\
 \hline
Our model & $300$ & $301$ & $350$ & $351$ & $0.0497$  & $0.070$  & $1.6\times10^{-11}$ & $1.7\times10^{-11}$&3.86  &2.09 \\
\hline
\end{tabular}
\caption{The couplings $\l_{L_1}~\rm{and}~\l_{L_2}$ are considered to be 0.01 with $\l_{12}=0$. }
\label{tab3}
\end{table}

\section{Conclusion}
\label{sec5}
While there is no compelling reason to believe the dark matter sector to be composed of only one type of particle, it still remains 
as a well motivated scenario particularly from minimality point of view. However, in certain minimal models, there can be motivations for a multi-component dark matter framework originating from inadequacies of a single component to fit for dark matter relic density as well as connections to other interesting observed phenomena, like neutrino mass. Motivated by this, we have studied a minimal extension of the Standard Model by two right handed neutrinos and two scalar doublets having non-trivial transformations under an unbroken $\mathbb{Z}_2 \times \mathbb{Z}'_2$ symmetry. While light neutrino masses arise at one loop level in a way similar to the scotogenic models consisting of particles from both the discrete $\mathbb{Z}_2$ sectors (playing the role of loop mediators), 
this minimal choice of the additional particles result in vanishing lightest neutrino mass. 

The two dark matter candidates are chosen to be the neutral components of the two scalar doublets which are thermally produced in the early universe by virtue of their electroweak gauge interactions. We particularly focus on the mass range in between $W$ boson mass and approximately 500 GeV where single component scalar doublet dark matter can not satisfy correct relic abundance. While scalar doublet relic does not depend crucially upon the Yukawa couplings to leptons, turning the Yukawa interactions on has interesting consequences for relic due to additional coannihilation channels. The inter-conversion between two dark matter candidates also play instrumental role in deciding the total relic abundance. We show that the model can satisfy correct total relic abundance criteria in the intermediate mass regime in agreement with the latest data from Planck mission. In addition to this, the model remains very much predictive at ongoing direct, indirect detection experiments as well as rare decay experiments looking for charged lepton flavour violation. Unlike typical multi-component dark matter models, our model gets more restrictive due to electroweak gauge interactions as well as the non-trivial roles both the dark matter particles play in generating light neutrino masses, opening up detection prospects through lepton portals like charged lepton flavour violation. While the origin of the discrete symmetries $\mathbb{Z}_2 \times \mathbb{Z}'_2$ remains unexplained in the current work, we leave a more detailed study looking for UV completion of such models to an upcoming work.

\acknowledgments
DB acknowledges the support from IIT Guwahati start-up grant (reference number: xPHYSUGI-ITG01152xxDB001), Early Career Research Award from DST-SERB, Government of India (reference number: ECR/2017/001873) and Associateship Programme of IUCAA, Pune. RR thank Biswajit Karmakar, Abhijit Kumar Saha and Amit Dutta Banik 
for various useful discusions while carrying out this work.

\end{document}